\documentclass[a4paper,11pt]{article}
\pdfoutput=1 

\usepackage{jcappub} 

\usepackage[T1]{fontenc} 
\usepackage{subfig}
\usepackage{physics}

\title{\boldmath CMB-PAInT: An inpainting tool 
for the cosmic microwave background}

\author[a, b]{C. Gimeno-Amo,}
\author[a]{E. Martínez-González,}
\author[a]{R. B. Barreiro}

\affiliation [a]{Instituto de Física de Cantabria, CSIC-Universidad de Cantabria,\newline Avda. de los Castros s/n, 39005 Santander, Spain.}
\affiliation[b]{Departamento de Física Moderna, Universidad de Cantabria,\newline Avda. de los Castros s/n, 39005 Santander, Spain.}

\emailAdd{gimenoc@ifca.unican.es}
\emailAdd{martinez@ifca.unican.es}
\emailAdd{barreiro@ifca.unican.es}

\abstract{The presence of astrophysical emissions in microwave observations forces us to perform component separation to extract the Cosmic Microwave Background (CMB) signal. However, even in the most optimistic cases, there are still strongly contaminated regions, such as the Galactic plane or those with emission from extragalactic point sources, which require the use of a mask. Since many CMB analyses, especially the ones working in harmonic space, need the whole sky map, it is crucial to develop a reliable inpainting algorithm that replaces the values of the excluded pixels by others statistically compatible with the rest of the sky. This is especially important when working with $Q$ and $U$ sky maps in order to obtain $E$- and $B$-mode maps which are free from $E$-to-$B$ leakage. In this work we study a method based on Gaussian Constrained Realizations (GCR), that can deal with both intensity and polarization.
Several tests have been performed to asses the validation of the method, including the study of the one-dimensional probability distribution function (1-PDF), E- and B-mode map reconstruction, and power spectra estimation. We have considered two scenarios for the input simulation: one case with only CMB signal and a second one including also Planck PR4 semi-realistic noise. Even if we are limited to low resolution maps, $N_{\mathrm{side}} = $ 64 if $T$, $Q$ and $U$ are considered, we believe that this is a useful approach to be applied to future missions such as LiteBIRD, where the target are the largest scales.}

\begin{document}
\maketitle
\flushbottom

\newpage

\section{Introduction} \label{Intro}

The anisotropies both in temperature and polarization of the \textit{Cosmic Microwave Background} (CMB) are one of the most important cosmological probes to test accurately the properties of the Lambda-Cold Dark Matter ($\Lambda$CDM) model \cite{2020A&A...641A...6P}. They allow us to study the largest scales and the earliest times of the Universe. According to the simplest slow-roll inflationary scenarios, the primordial fluctuations, the seeds for late universe structures, are a single realization of an isotropic Gaussian random field \cite{2020A&A...641A..10P}. Consequently all the statistical information is encapsulated in the temperature and polarization angular power spectrum, making crucial their correct estimation.

One of the main issues to characterize the CMB anisotropies is how to deal with extragalactic and Galactic foreground emissions such as synchrotron, free-free, or thermal dust. These are astrophysical emissions between the last scattering surface and us. Precisely, the CMB has been measured in several frequency channels to exploit their different frequency response and to allow component-separation algorithms \cite{Planck:2018yye, 2012MNRAS.420.2162F, 2008ApJ...676...10E, 2008arXiv0803.1814C, 10.1111/j.1365-2966.2011.19770.x} to subtract an important part of them. Unfortunately, strongly contaminated regions such as the Galactic plane and the locations of extragalactic point sources can not be used for the statistical analyses, even after applying component-separation methods. In these regions, foregrounds are too large and complicated to be fully cleaned. The standard approach is to mask these regions using the confidence mask provided by each method, but this leads to loss of information and other technical difficulties. For instance, this prevents us from accurate characterization of large angular scales, which is crucial for the study of CMB anomalies \cite{2004ApJ...605...14E,2004ApJ...609...22V,2004PhRvD..69f3516D, 2004MNRAS.354..641H, 2006A&A...454..409B, 2010AdAst2010E..77V, 2009ApJ...704.1448H, Planck:2019evm, Gimeno-Amo_2023}. In polarization the mask is even more problematic, not just at the harmonic level, but also at the map level. Masking $Q$ and $U$ Stokes parameters introduces an undesired mixing between $E$- and $B$-modes \cite{PhysRevD.64.063001, Lewis:2003an}, also referred as $E$-to-$B$ leakage. Actually, this is critical for the detection of the primordial $B$-mode signal as the $E$-mode signal is significantly larger. At map level, the lack of full-sky $Q$ and $U$ maps prevents us from having an accurate $E$- and $B$-mode reconstruction. The reason for this is the non-local nature of the transformations between $Q$ and $U$, and $E$ and $B$. Indeed, the main motivation of this work is an accurate reconstruction of the $E$- and $B$-mode outside the mask. These reconstructed maps are of interest for studies of isotropy and statistics, including the anomalies presented in the large angular scales. In order to characterise the quality of the inpainted maps, a few tests have been used including the power spectra. However, we note that for estimating the power spectra in masked skies other standard techniques are available which are described in the next paragraph.

Given this situation, the possible solutions are: (1) adapt the estimators to partial sky maps, which in general are more complicated and numerically expensive, or (2) fill-in the maps using the so-called inpainting algorithms. For instance, there are several methods to recover the angular power spectra from partial sky coverage. In \cite{Alonso:2018jzx} a very fast algorithm is described based on the pseudo-$\mathcal{C}_{\ell}$ formalism to estimate the angular power and cross-power spectra. This is the NaMaster implementation, and it works for spin-0 and spin-2 fields, which is able to correct for the coupling between multipoles induced by the mask and to purify $E$- and $B$-modes. However, it is sub-optimal for the large scales. Quadratic Maximum Likelihood (QML) \cite{Tegmark:1996qt, PhysRevD.64.063001, Bilbao-Ahedo:2021jhn} is an alternative method, which is optimal for large angular scales but computationally demanding. On the other hand, inpainting techniques aim to reconstruct a full-sky CMB map statistically coherent with the observed region. Inpainting techniques are not novel, they were already used in the image processing to fill in missing pixels and restore blurred photographs \cite{999016}. This idea was extended to the Cosmology field, and in particular, it has been applied to fill the CMB \cite{2008StMet...5..289A, 2008PhRvD..77l3539I, 2009arXiv0903.1308P, 2012MNRAS.424.1694B, 2012ApJ...750L...9K}. The simplest approach is a diffuse inpainting. The algorithm iteratively fills the masked pixels by averaging the closest neighbours. There are more sophisticated methodologies such as the purified inpainting followed in \cite{Planck:2019evm}. These methodologies do not assume any underlying model. In this work we consider an alternative approach based on Gaussian constrained realizations \cite{1991ApJ...380L...5H, 2012MNRAS.424.1694B, 2012ApJ...750L...9K, 2013A&A...555A..37B, Marcos-Caballero:2019jqj}. Assuming the Gaussianity of the field, we take the power spectra that best explain the data and use it to get the correlations between pixels. Then, we fill the missing pixels sampling from the conditional probability distribution. The exact approach is not feasible for full-resolution maps, with millions of pixels. However, it can be useful for low resolution maps, up to $N_{\mathrm{side}}$ = 64 and $\ell_{\mathrm{max}}$ = 192, covering the full range of the reionization and the recombination peaks of the $B$-mode. In particular, it could be a powerful tool for future missions such as LiteBIRD \cite{LiteBIRD:2022cnt}, which aims to measure large angular scales in polarization. Recently, the use of Neural Networks has been proposed as an alternative way to inpaint the CMB \cite{Puglisi:2020deh}. 

The methodology presented in this work and, in particular, the Python code have been successfully used in \cite{Gimeno-Amo_2023} to reconstruct accurately the $E$-mode map where hemispherical power asymmetry was studied. 

The outline of this paper is as follows. In section \ref{GCR}, we present the inpainting methodology. In section \ref{Sims}, we describe briefly the simulations used in this work. In section \ref{Test_Validation}, we apply our inpainting technique to simulated data and present the results for some tests, including the power spectra and $E$-mode map reconstruction. In section \ref{PythonCode}, a Python implementation of the methodology is presented, the \texttt{CMB-PAInT} package. In section \ref{Conclusions}, we summarize our work.

\section{Gaussian Constrained Realization} \label{GCR}

The inpainting technique presented in this work is a pixel domain approach based on a Gaussian Constrained Realization (GCR) \cite{2012MNRAS.424.1694B, 2012ApJ...750L...9K}. The followed methodology was already described in \cite{Marcos-Caballero:2019jqj} for the temperature field, but in this work we extend it to a spin-2 field, i.e., to CMB polarization. The idea is to fill the masked pixels by sampling from the conditional probability distribution, $p(\mathbf{\hat{d}}|\mathbf{d})$, where $\hat{d}$ is the vector of the inpainted field and d is the vector of the available pixels. The method requires a single assumption, Gaussianity of the field, which is a good approximation for the CMB data. Under this condition, only the pixel covariance matrix is needed, which can be computed given a theoretical power spectrum\footnote{In the present work we have used as the input power spectra the best fit to the $\Lambda$CDM mode provided by Planck \cite{2020A&A...641A...5P}. It can be downloaded from the Planck Legacy Archive: \url{pla.esac.esa.int}}, $C_{l}$, following the next equations (see Appendix A of \cite{PhysRevD.64.063001} for the full set of equations):
\begin{equation} \label{eq:1}
    C_{ij}^{TT} = \langle T_{i}T_{j} \rangle = \sum_{\ell}\left(\frac{2\ell+1}{4\pi}\right)C_{\ell}^{TT}P_{\ell}(z)
\end{equation}

\begin{equation} \label{eq:2}
    C_{ij}^{QQ} = \langle Q_{i}Q_{j} \rangle = \sum_{\ell} \left(\frac{2\ell +1}{4\pi}\right)[F_{\ell}^{12}(z)C_{\ell}^{EE}-F_{\ell}^{22}(z)C_{\ell}^{BB}]
\end{equation}

\begin{equation} \label{eq:3}
    C_{ij}^{UU} = \langle U_{i}U_{j} \rangle = \sum_{\ell} \left(\frac{2\ell +1}{4\pi}\right)[F_{\ell}^{12}(z)C_{\ell}^{BB}-F_{\ell}^{22}(z)C_{\ell}^{EE}]
\end{equation}

\begin{equation} \label{eq:4}
    C_{ij}^{TQ} = \langle T_{i}Q_{j} \rangle = -\sum_{\ell} \left(\frac{2\ell +1}{4\pi}\right)F_{\ell}^{10}(z)C_{\ell}^{TE}
\end{equation}
where $z = \hat{r_{i}}\cdot \hat{r_{j}}$ gives the cosine of the angle between the two pixels. The $F$ functions are 

\begin{equation}
    F^{10}(z) = 2\frac{\frac{\ell z}{(1-\ell^{2})}P_{\ell-1}(z)-\left(\frac{\ell}{1-z^{2}}+\frac{\ell(\ell-1)}{2}\right)P_{\ell}(z)}{[(\ell-1)\ell(\ell+1)(\ell+2)]^{1/2}}
\end{equation}

\begin{equation}
    F^{12}(z) = 2\frac{\frac{(\ell+2)z}{(1-\ell^{2})}P_{\ell-1}^{2}(z)-\left(\frac{\ell-4}{1-z^{2}}+\frac{\ell(\ell-1)}{2}\right)P_{\ell}^{2}(z)}{(\ell-1)\ell(\ell+1)(\ell+2)}
\end{equation}

\begin{equation}
    F^{22}(z) = 4\frac{(\ell+2)P_{\ell-1}^{2}(z)-(\ell-1)zP_{\ell}^{2}(z)}{(\ell-1)\ell(\ell+1)(\ell+2)(1-z^{2})}
\end{equation}
where $P_{\ell}$ are the Legendre polynomials.

The covariance matrix will be arranged by blocks as follows:
\begin{equation}
{M}(\hat{r}_{i}\cdot \hat{r}_{j}) = 
\begin{pmatrix}
\langle {T}_{i}{T}_{j} \rangle & \langle {T}_{i}{Q}_{j} \rangle & \langle {T}_{i}{U}_{j} \rangle\\
\langle {T}_{i}{Q}_{j} \rangle & \langle {Q}_{i}{Q}_{j} \rangle & \langle {Q}_{i}{U}_{j} \rangle\\
\langle {T}_{i}{U}_{j} \rangle & \langle {Q}_{i}{U}_{j} \rangle & \langle {U}_{i}{U}_{j} \rangle
\end{pmatrix} 
\end{equation}
Note that the elements of this covariance matrix are referred to a specific coordinate system where the reference direction points along the great circle connecting the two points. However, what is needed is the covariance matrix referred to a global coordinate frame where the reference directions are meridians, so the following rotation matrices are applied (see Appendix A in \cite{PhysRevD.64.063001} to find how to compute the rotation angle\footnote{For those who want to use the equations, there is a small typo in equation A7 of the appendix since the proportionality constant is negative, not positive. Thus in equation A8, the negative sign corresponds to the case where the z component of the vector $\hat{r}_{ij}$, the vector of the great circle connecting the two pixels, is positive, and vice versa.} $\alpha$):

\begin{equation}\label{Eq.2.9}
    \langle {x}_{i}{x}_{j}^{t} \rangle = {R}(\alpha_{ij}){M}({\hat{r}}_{i}\cdot {\hat{r}}_{j}){R}(\alpha_{ij})^{t}
\end{equation}

\begin{equation}\label{Eq.2.10}
    \mathbf{R}(\alpha) = 
\begin{pmatrix}
1 & 0 & 0\\
0 & \cos{2\alpha} & \sin{2\alpha}\\
0 & -\sin{2\alpha} & \cos{2\alpha} 
\end{pmatrix} 
\end{equation}
where $x_{i}$ = $\{T, Q, U\}$.

The effect of the experimental beam and the pixel window function can be added by smoothing the power spectra in equations (\ref{eq:1})-(\ref{eq:4}). 

The maximum multipole considered in the sum of equations (\ref{eq:1})-(\ref{eq:4}) should be equal to the largest multipole accounted for in the map. As default $3N_{\mathrm{side}}$ is considered. This approach is only feasible for low resolution maps, up to $N_{\mathrm{side}} = 64$, as the dimension of the total matrix is $3N_{\mathrm{pix}} \times 3N_{\mathrm{pix}}$. Adding a small regularizing noise to the diagonal is needed to avoid 
singularities and to ensure that the matrix is positive definite. The noise level depends on the resolution. For $N_{\mathrm{side}} = 64$ a noise amplitude of 0.00001\% is used. A similar regularization can be achieved by considering in the sum 
a maximum multipole of 4$N_{\mathrm{side}}$ or even larger. For a detailed discussion on the regularity of a CMB covariance matrix see \cite{2017JCAP...02..022B}.

Once the matrix is computed and rotated, we reorder the columns and rows in a way that all the unmasked pixels are in the first entries and the masked ones in the last entries\footnote{In the case of a full $TQU$ covariance matrix, the order considered is: unmasked $T$ pixels, unmasked $Q$ pixels, unmasked $U$ pixels, masked $T$ pixels, masked $Q$ pixels and masked $U$ pixels.}. Then, the Cholesky decomposition allows one to sample from the desired distribution by solving the following system,
\begin{equation}
\label{chol}
\begin{pmatrix}
\mathbf{d}\\
\mathbf{\hat{d}}
\end{pmatrix} = 
\begin{pmatrix}
\mathbf{L} & 0\\
\mathbf{R} & \mathbf{\hat{L}}
\end{pmatrix} 
\begin{pmatrix}
\mathbf{z}\\
\mathbf{\hat{z}}
\end{pmatrix} 
\end{equation}
The matrix in the right-hand side is the Cholesky decomposition, where L and $\hat{\mathrm{L}}$ are low triangular matrices and R a rectangular matrix. The number of rows and columns of the L matrix is equal to the number of unmasked pixels, while for $\hat{\mathrm{L}}$ is the number of masked pixels.

Looking at equation \ref{chol} and taking into account that $\mathbf{L}$ is a lower triangular matrix, it becomes apparent that a matrix inversion is not needed and the vector $\mathbf{z}$ can be computed in a recursive way, 
\begin{equation}
    \label{Eq:2.12}
    z_{n} = \frac{d_{n}-\sum_{k=1}^{n-1}L_{nk}z_{k}}{L_{nn}}
\end{equation}
If the model is coherent with the observed data this vector should be a Gaussian random vector with zero mean and unit variance. Then, a new random vector $\mathbf{\hat{z}}$, also following a $\mathcal{N}(0, 1)$ distribution, is generated, and the field $\mathbf{\hat{d}}$ is sampled, which has two contributions: the constrained part and the unconstrained or stochastic part.
\begin{equation}
    \label{Eq:2.13}
    \mathbf{\hat{d}} = \mathbf{Rz}+\mathbf{\hat{L}\hat{z}}
\end{equation}

In this procedure, inpainting is performed simultaneously on the $TQU$ maps. 
Of course, if we are only interested in the temperature map, we can just compute the $TT$ block, which will significantly reduce the computational cost and similarly if one is only interested in polarization.
However, if the $TT$, $TQ$, and $TU$ blocks are not included, the inpainted pixels will not have the correlation between temperature and polarization (with comes from $TE$ correlation in the standard cosmology, or from $TE$, $TB$ and $EB$ in beyond $\Lambda CDM$ models). In that scenario the size of the matrix will be reduced to $4/9$ if we just take into account $Q$ and $U$ and to $1/9$ if we just want $T$. 
Even if the correlation between $TE$ is low, the $Q$ and $U$ pixels located outside the polarization mask will further restrict the potential values of the temperature pixels within the mask, and vice versa for the polarization pixels inside the mask.

In principle, the covariance matrix should include all the components present in the map to be inpainted. Dealing with the Planck polarization data, the dominant component is the noise and systematics, and their contribution must be considered. If not, mismatches between the pixels outside the mask and the covariance matrix introduces artifacts in the map. As there is not any theoretical model for the noise and systematics, the only way to estimate the covariance matrix is from end-to-end (E2E) simulations. This can be a limitation of the method: the number of simulations needed for a good characterization of the realistic anisotropic and correlated noise and systematics, is at least of a few thousands (see Appendix \ref{Appendix_A}), while the available realistic realizations are usually limited to several hundreds. 

\section{Signal and noise simulation} \label{Sims}

In order to validate our method, we generate a single Gaussian isotropic CMB map at a resolution of $N_{\mathrm{side}} = 64$. We use the healpy\footnote{\url{ https://github.com/healpy/healpy}} function \texttt{synfast}. We smooth
the power spectra with a Gaussian beam of $160^\prime$, and we consider a maximum multipole $\ell_{max} = 3N_{\mathrm{side}}$. We generate 1200 inpainting realizations from a single sky realization based on the PR3 $\Lambda$CDM best fit model. The same spectra is used to compute the pixel covariance matrix. Planck 2018 temperature and polarization confidence masks define the region to be filled. At $N_{\mathrm{nside}}$ = 64 they respectively leave 71.3\% and 72.4\% of the sky available. 

In order to validate the method in the presence of correlated and anisotropic noise, we generate a semi-realistic noise simulation using characteristics from Planck, an ESA satellite that observed the CMB over the full-sky with an unprecedented sensitivity and frequency coverage \cite{2011A&A...536A...1P}. We start by computing the covariance matrix from the 600 end-to-end (E2E) Planck Release 4 (PR4) \cite{2020A&A...643A..42P} noise simulations, that include also the expected systematics, except foreground residuals, propagated through the Sevem component separation pipeline \cite{2012MNRAS.420.2162F}. Then, we generate a Gaussian random realization with the proper correlations given by the E2E simulations\footnote{Actually, this is also done using a Cholesky decomposition. In this case, we obtain the realization as $\mathbf{d^\prime} = \mathbf{L^\prime} \mathbf{z^\prime}$, where $\mathbf{L^\prime}$ is the Cholesky decomposition of the covariance matrix and $\mathbf{z^\prime}$ a Gaussian random vector with zero mean and unity dispersion.}. Following this pixel-based approach we are able to simulate not only the correlations but also the anisotropy. However, it does not include the presence of possible non-Gaussianity generated by systematic effects.

\begin{figure}[t!]
    \centering
    \includegraphics[scale = 0.38]{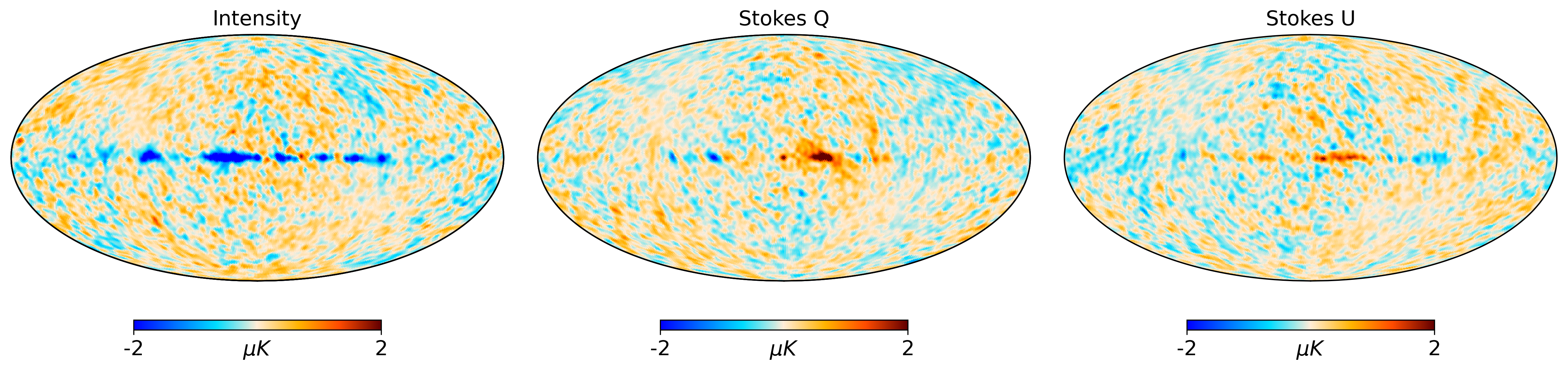}
    \caption{Input $TQU$ semi-realistic noise simulation at $N_{\mathrm{side}}$ = 64 and convolved with a Gaussian beam of FWHM=$160^\prime$.}
    \label{fig:Semi-Realistic}
\end{figure}

Figure \ref{fig:Semi-Realistic} shows the input semi-realistic noise realization. The polarization noise level used in this paper is very large compared to the noise expected for the future experiments, but it mimics the level of instrumental noise and systematics present in the Planck foreground-cleaned CMB Sevem map for PR4.

\section{Tests and Validation} \label{Test_Validation}

In this section, a series of tests are conducted on a set of inpainted realizations, derived from a single input sky, to assess the algorithm's performance. Planck 2018 temperature and polarization confidence masks \cite{Planck:2018yye} are considered to inpaint the input maps. First, we verify that the mean and variance maps are compatible with the expected values. Then, some statistics are examined, such as the 1-point probability density function (1-PDF) in real space or the power spectrum in harmonic space confirming their consistency with the input values. Furthermore, for each inpainted $Q$ and $U$ maps, $E$- and $B$-mode maps are generated. By subtracting them from the corresponding input maps, we compute the mean and standard deviation of the residuals pixel by pixel. This provides information on how the residuals are distributed and on the level of the errors introduced by the inpainting.
In particular, this map can be used as a suitable reference to generate customized $E$- and $B$-mode masks for pixel based estimators.

All of these tests are applied in two different scenarios: the noiseless case where the input sky is just CMB, or the case of CMB plus a semi-realistic noise simulation (constructed as explained in section \ref{Sims}). The second scenario can also be divided into some subcases depending on how well the pixel-pixel covariance matrix is characterized. We first study the ideal case where the correlations between pixels are well known and covariances are perfectly characterized. This is possible because the semi-realistic noise simulations are generated from an input covariance matrix. Thus, the performance of the inpainting in the presence of correlated and anisotropic noise in an ideal case can be studied. However, as previously mentioned, in a real experiment we do not expect to have a perfect knowledge of the complex properties of instrumental noise and systematics. The covariance matrix will need to be estimated from high-cost CPU simulations, limiting the number that can be produced (typically only several hundreds) and, therefore, our capacity to characterize it properly.
In Appendix \ref{Appendix_A} we study the impact of the matrix convergence varying the number of simulations used to estimate the covariance matrix. In particular, this can be used to establish a rough estimation of the number of E2E simulations needed from future experiments. To study the convergence of the matrix the differences between the input and estimated matrices can be checked, but other variables can also be studied such as the intermediate $z$ variables (see eq. \ref{Eq:2.12}). In presence of a mismatch between the covariance matrix and that of the unmasked pixels, an error is produced which propagates in the calculation of the $z$ variables. Thus, these quantites are not longer $\mathcal{N}$(0,1) variables as the dispersion increases, and it can be used as a tracer of the convergence. Additionally, the constrained part of the inpainting (first term in the right-hand side of Eq. \ref{Eq:2.12}) is also affected, and this introduces artifacts within the inpainted region. In any case, the number of simulations needed for a good convergence inferred from the study of Appendix \ref{Appendix_A} should be taken as a tentative number. For the realistic E2E simulations there are other effects that can contribute to the mismatch such as the non-Gaussianities, which are not simulated here. 

Nevertheless, in the most ideal case we would be interested in no recovering the prominent noise and systematics in the inpainted region, but holding the statistical compatibility with the unmasked pixels. However, taking into account that not including them in the methodology generates some artifacts, there is not straightforward way to proceed. Fortunately, there is a situation where this can be avoided. If the noise is negligible compared to the signal at the map level, which is the case for the Planck temperature maps, its contribution can be also neglected from the matrix without having to pay the penalty of a significant mismatch. For the most general case there are other alternatives that we will leave for further studies. For example, noise and systematics can be isotropized for all the matrix elements that involve masked pixels, or directly their contribution can be removed and just take into account the noise and systematics for the unmasked pixels, where they are expected to be subdominant. The last option could lead to matrix singularity problems, which will need the inclusion of regularization noise to be solved. In any case, the optimal solution will depend on the nature of the data and the estimator to be used. For instance, in \cite{Gimeno-Amo_2023} we show that performing inpainting using a noise covariance matrix estimated from a set of 300 simulations was sufficient to improve significantly the performance of our estimator with respect to a simple masking approach in Q and U.

\subsection{Constrained contribution}

In this section we study the mean map within the inpainted region by averaging over the $1200$ realizations, and we compare it with the theoretical prediction. We also show an example of inpainted maps, including the constrained and unconstrained contributions. We consider both scenarios, only CMB and adding semi-realistic noise.

From eq.~\ref{Eq:2.13} it is straightforward to obtain that the mean and covariance of the inpainted field are, 
\begin{equation}
    \expval{\mathbf{\hat{d}}} = \mathbf{Rz} = \mathbf{R(L^{-1}d)}
\end{equation}
\begin{equation}\label{C_matrix}
    \mathbf{\hat{C}} = \expval{\mathbf{\hat{d}\hat{d}^{t}}}-\expval{\mathbf{\hat{d}}}\expval{\mathbf{\hat{d}^{t}}} = \mathbf{\hat{L}\hat{L}^{t}}
\end{equation}

The mean map is given by the constrained part and it is the dominant contribution to the inpainting in the regions close to the boundaries of the mask, where the constraints are tighter. The variance in these regions is close to zero, as these pixels are almost fully constrained and their value do not vary significantly from one inpainting realization to another.

\begin{figure}[t!]
    \centering
    \includegraphics[scale = 0.4]{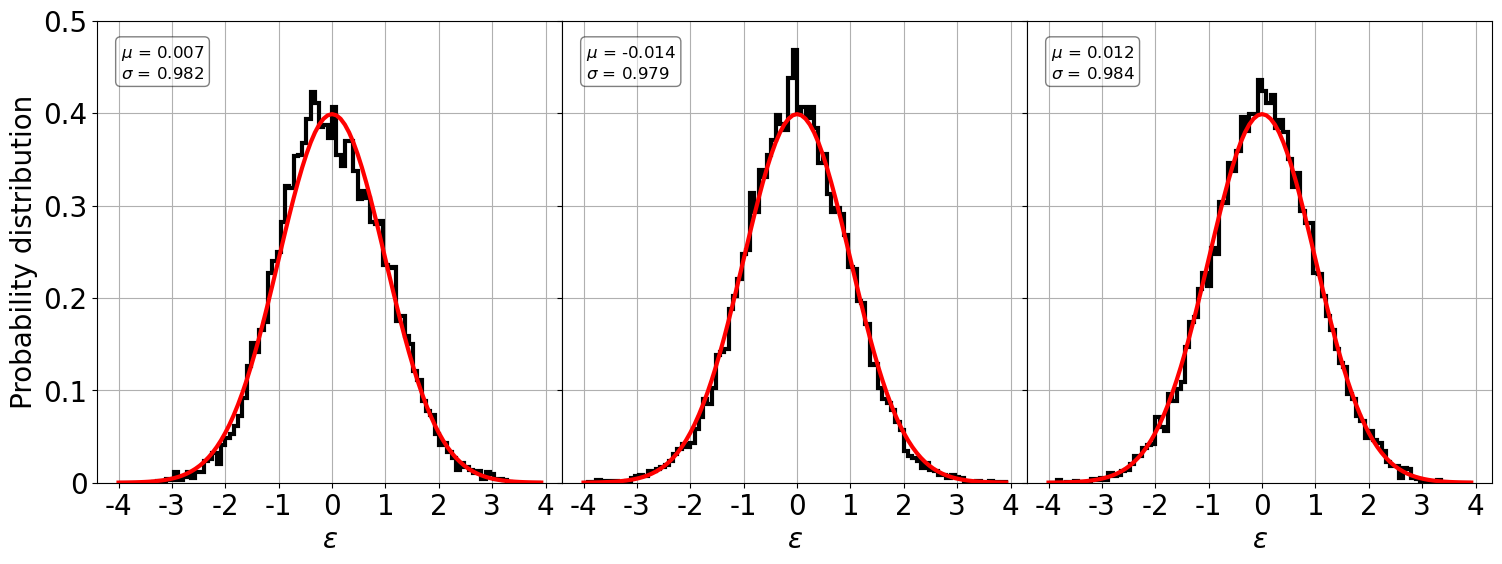}
    \caption{Distribution of $\epsilon$ variable for inpainted $T$, $Q$, and $U$ maps. For comparison, the $\mathcal{N}(0,1)$ distribution is also given (red line).}
    \label{Fig_Epsilon}
\end{figure}

In order to check if the mean field of the 1200 inpainted realizations ($N_{\mathrm{sims}}$) is consistent with the theoretical prediction, we define the quantity
\begin{equation}\label{epsilon}
    \mathbf{\epsilon} = \frac{\expval{\mathbf{\hat{d}}}_{\mathrm{obs}}-\expval{\mathbf{\hat{d}}}_{\mathrm{th}}}{\sigma\left(\expval{\mathbf{\hat{d}}}_{\mathrm{obs}}\right)/\sqrt{N_{\mathrm{sims}}}}
\end{equation}
which measures pixel by pixel if the observed difference is compatible with 0 given the expected error. If everything is consistent, $\epsilon$ should follow a Gaussian distribution with zero mean and unit variance, $\mathcal{N}(0, 1)$. 
This is actually seen in figure \ref{Fig_Epsilon}, where we show the results for the $T$, $Q$ and $U$ components for the noiseless scenario. Similar results are obtained when the case including semi-realistic noise is considered, showing that the method also works when a well characterised anisotropic and correlated noise is added to the input CMB sky.

\begin{figure}[t!]
    \centering
    \includegraphics[scale = 0.44]{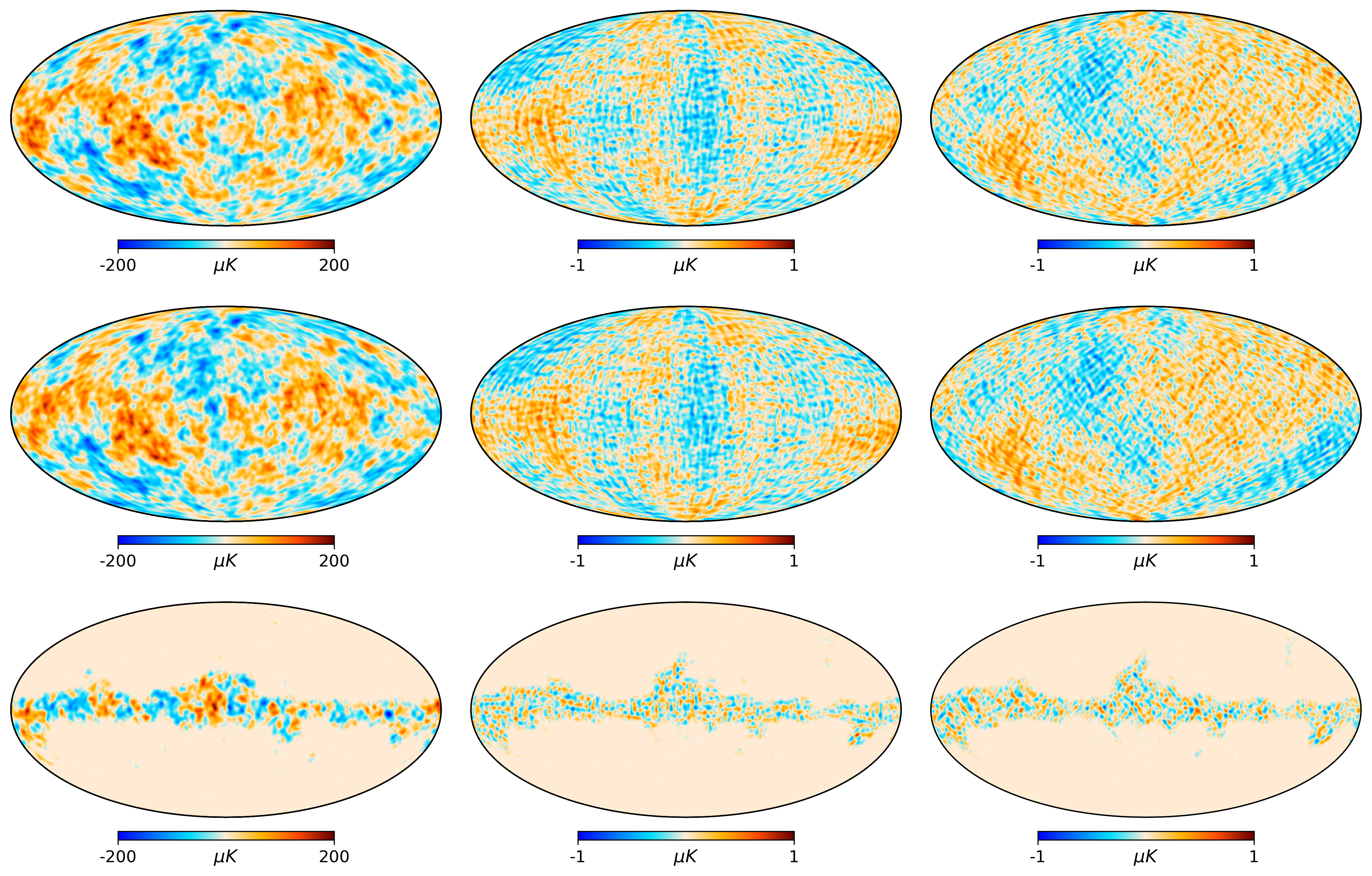}
    \caption{Example of an inpainted realization in the case where only the CMB signal is considered. First row corresponds to input $T$ (first column), $Q$ (second), and $U$ (third) maps. An inpainted realization is shown in the second row, while the third one shows the difference. All the maps are at $N_{\mathrm{side}}$ = 64 and have a resolution of 
    $160^\prime$.}
    \label{Fig_Signal_Scenario}
\end{figure}

\begin{figure}[t!]
    \centering
    \includegraphics[scale = 0.44]{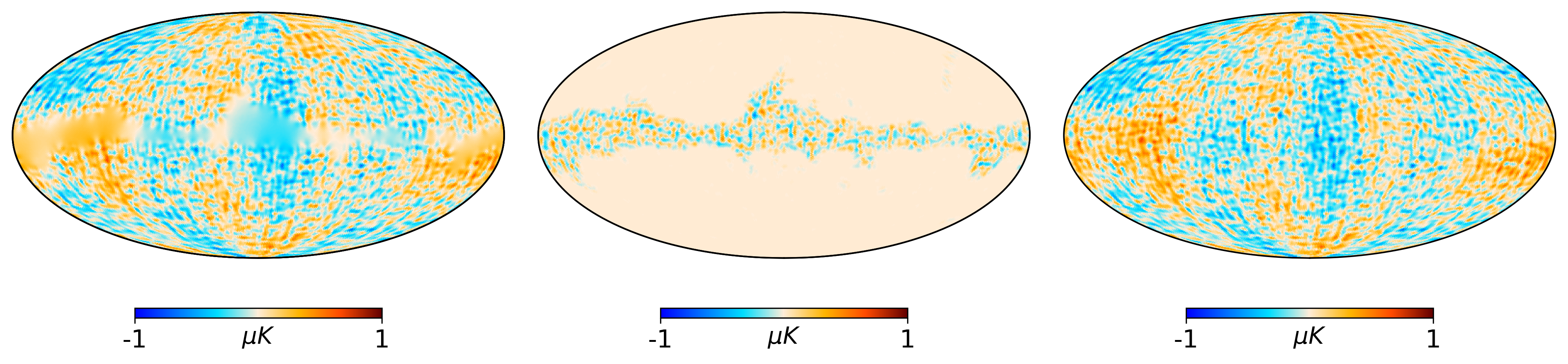}
    \caption{Example of an inpainted realization for the $Q$ Stokes parameter in the case where only the CMB signal is considered. Left and middle panels show the constrained (deterministic) and unconstrained (stochastic) parts, while the right panel is the sum of both.}
    \label{Fig_Cons_Uncons}
\end{figure}

\begin{figure}[t!]
    \centering
    \includegraphics[scale = 0.44]{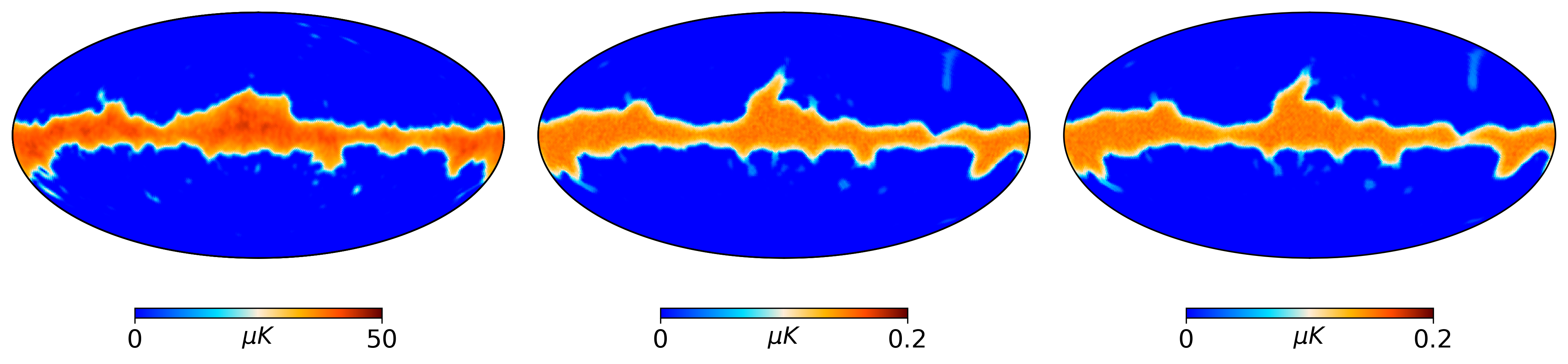}
    \caption{Dispersion of the inpainted maps estimated from the full set of 1200 inpainted realizations for the T (left panel), Q (middle), and U (right) components.}
    \label{Fig_Dispersion}
\end{figure}

Figure \ref{Fig_Signal_Scenario} shows an example of one inpainted realization for the noiseless scenario: the input (top panels), inpainted (middle) and difference (bottom) maps are shown for the $T$ (left column), $Q$ (middle) and $U$ (right) components. An example of the constrained and unconstrained components is showed in figure \ref{Fig_Cons_Uncons} for the $Q$ component, while Figure \ref{Fig_Dispersion} provides the dispersion maps for the three components. There is a clear gradient pointing from the central regions of the mask towards the boundaries where the variance tends to zero. As mentioned before, this is because the pixels in the boundary regions are strongly constrained.

\begin{figure}[t!]
    \centering
    \includegraphics[scale = 0.44]{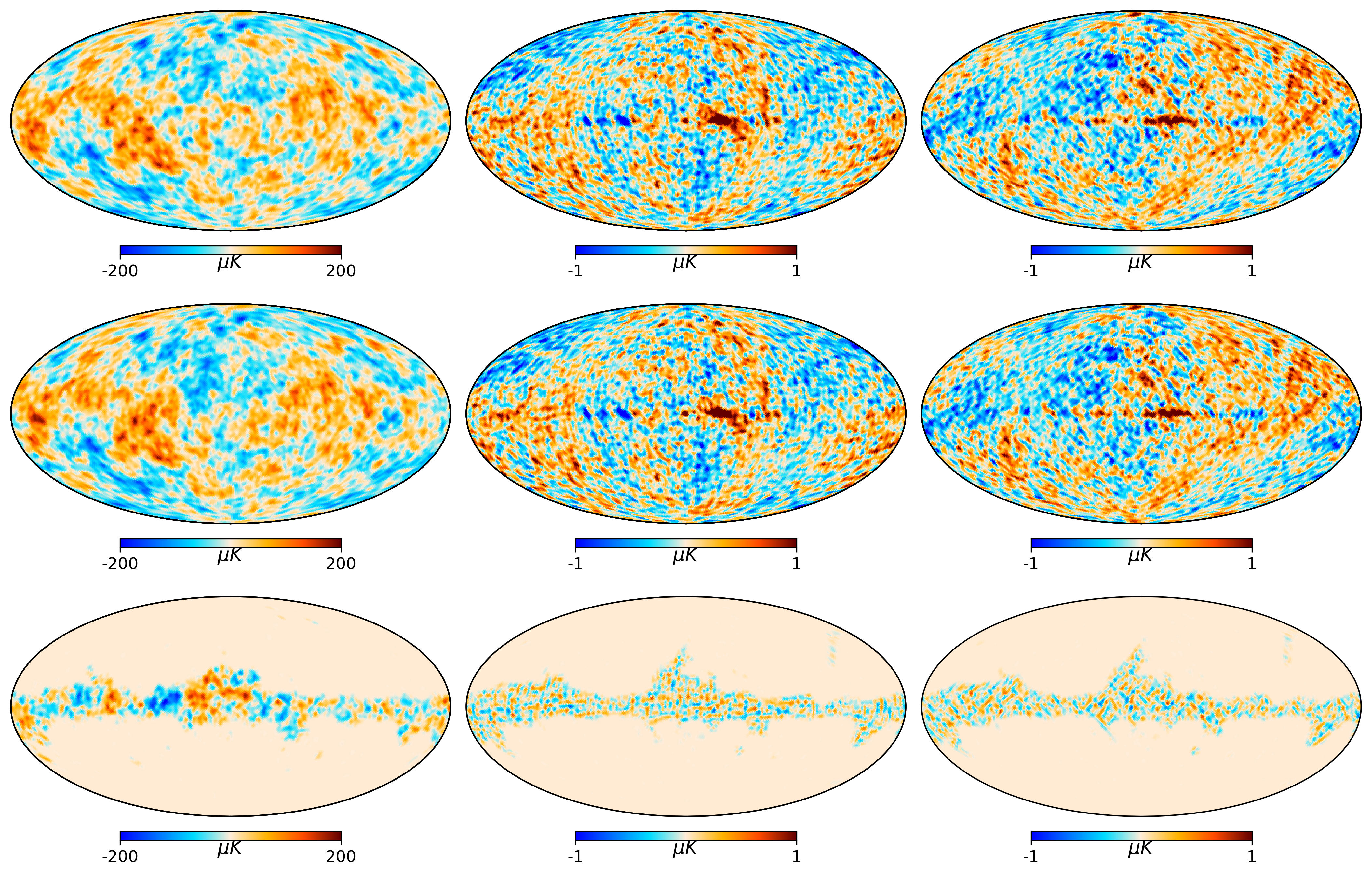}
    \caption{Example of an inpainted realization in the case where semi-realistic noise is added to the CMB signal. First row corresponds to input $T$ (first column), $Q$ (second), and $U$ (third) maps. An inpainted realization is shown in the second row while the third row shows the difference. All the maps are at $N_{\mathrm{side}}$ = 64 and have a resolution of $160^\prime$.}
    \label{Fig_Noise_Scenario}
\end{figure}

Additionally, an example for the scenario with noise is given in figure \ref{Fig_Noise_Scenario}. As seen, some bright anisotropic features of the Galactic plane are actually reproduced in the inpainted maps. This is due to the fact that a perfectly characterized anisotropic covariance matrix has much more information than an isotropic one, and therefore, the sampled values are more constrained.

\subsection{1-point probability distribution function}

\begin{figure}[t!]
    \centering
    \includegraphics[scale = 0.4]{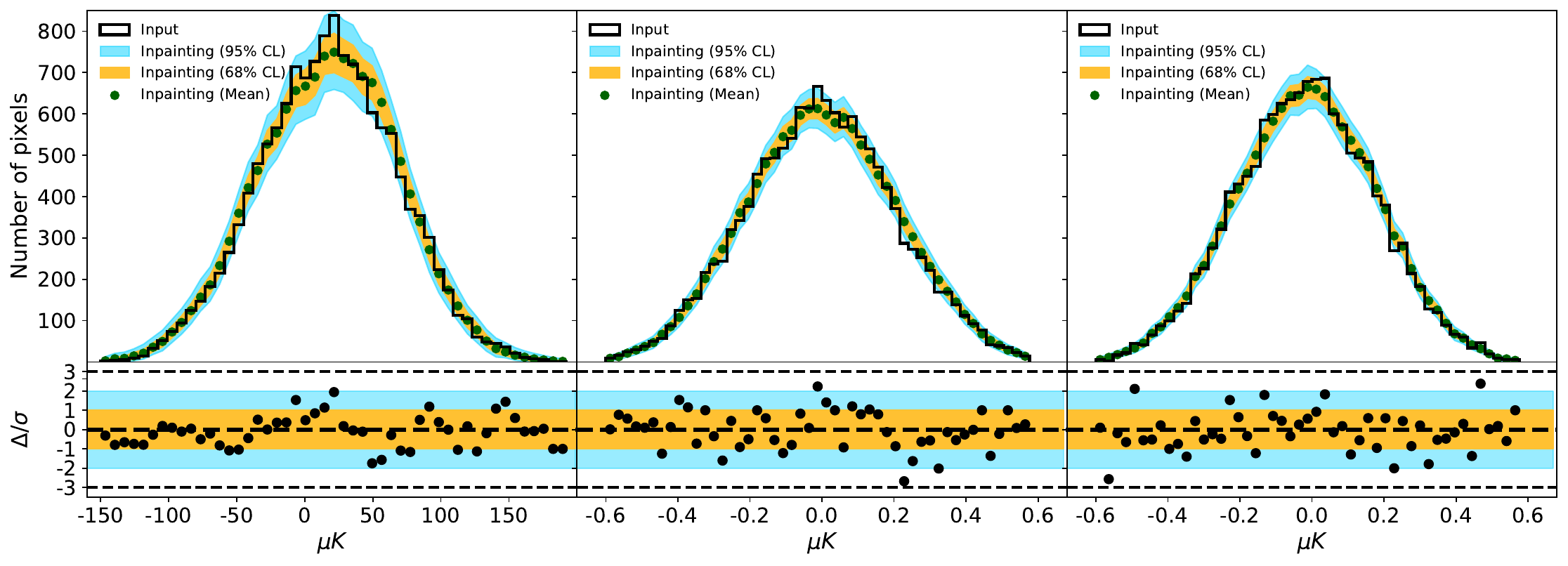}
    \caption{One-dimensional probability distribution inside the inpainted region for the $T$ (left panel), $Q$ (middle), and $U$ (right) components considering only the CMB signal. The black histogram corresponds to the input map. Green dots are the average value per bin obtained from the 1200 inpainted realizations. Orange and blue contours are the 68\% and 95\% C.L., respectively, obtained from the distribution of the inpainted maps. Residuals are also shown in the lower panel.}
    \label{1PDF_TQU}
\end{figure}

\begin{figure}[t!]
    \centering
    \includegraphics[scale = 0.4]{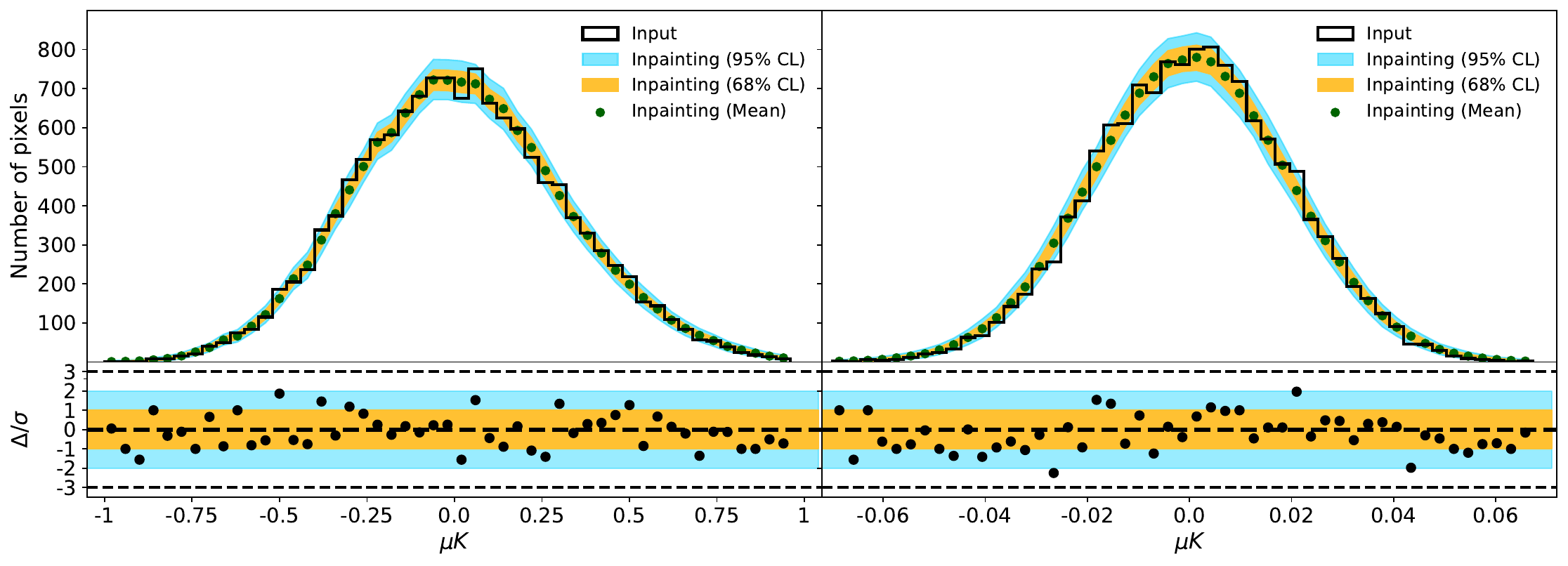}
    \caption{Same as figure \ref{1PDF_TQU}, but for $E$- and $B$-modes.}
    \label{1PDF_EB}
\end{figure}

\begin{figure}[t!]
    \centering
    \includegraphics[scale = 0.4]{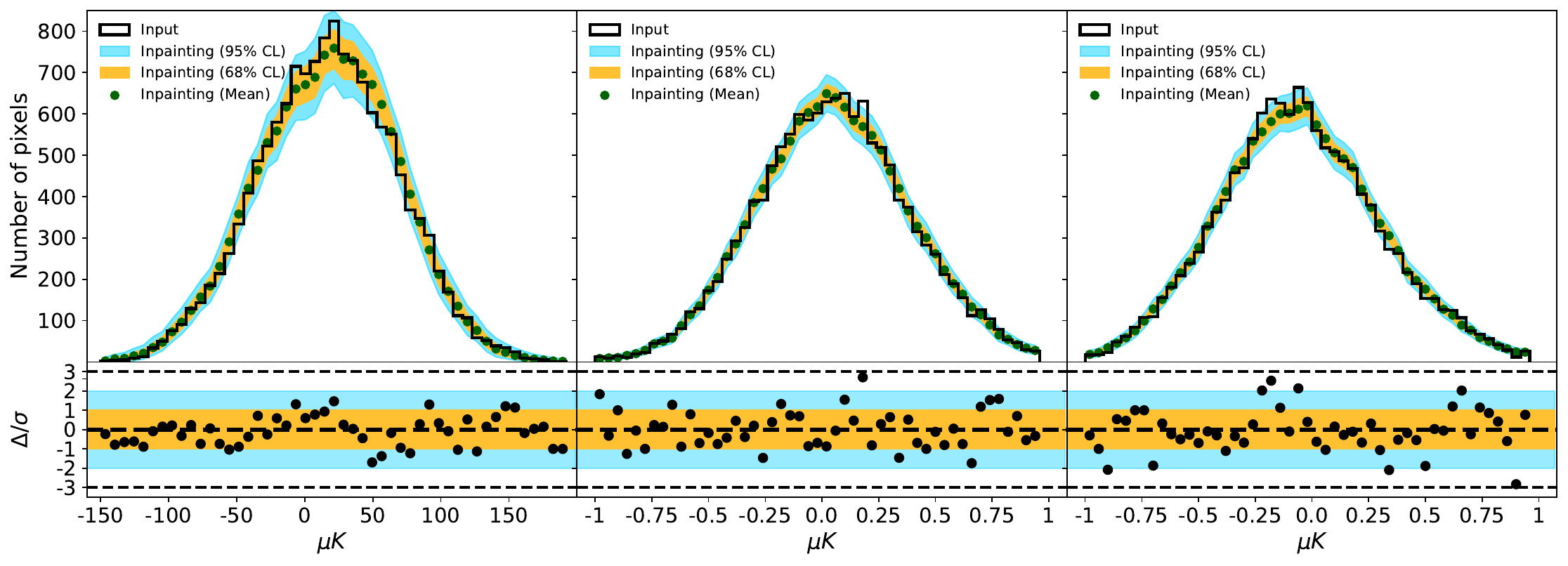}
    \caption{Same as figure \ref{1PDF_TQU}, but for the scenario where a semi-realistic noise realization is added to the CMB signal.}
    \label{TQU_Noise}
\end{figure}

In this section we study the 1-point probability distribution function (1-PDF) of the inpainted pixels. We also compute the 1-PDF for the $E$- and $B$-modes. Figures \ref{1PDF_TQU} and \ref{1PDF_EB} provide the PDFs for the $T$, $Q$, and $U$ components and the $E$- and $B$-modes, respectively, in the noiseless scenario. Similar results are obtained for the semi-realistic noise case, as shown in figure \ref{TQU_Noise}.

Inpainting performs well in both scenarios from the point of view of the reconstructed 1-PDF. The difference in each bin between the input value and the average over the 1200 inpainting realizations is within 2$\sigma$ for almost every point. For the semi-realistic noise scenario, the tails in the $Q$ and $U$ 1-PDF are larger due to the presence of noise and systematics.

\subsection{$E$- \& $B$-mode reconstruction}\label{EB_recon}

The quality of the recovery of the $E$- and $B$-mode maps is another crucial test. The transformation from $Q$ and $U$ Stokes parameters to more suitable variables $E$- and $B$-modes is not local. This means that full-sky $Q$ and $U$ measurements are needed in order to have accurate $E$- and $B$-mode maps free from $E$-to-$B$ leakage. Moreover, the reduction of this leakage is one of the main motivations of this work. Precisely, inpainting can fill the masked $Q$ and $U$ regions with a signal statistically compatible with the clean sky outside the mask, removing the potential foreground residuals. This approach is particularly useful in the case of pixel-based estimators that deal with the E- and B-modes maps. For harmonic-based estimators, there are alternative methodologies to deal with a mask. For instance to recover the CMB polarization power spectra at large scale, the Quadratic Maximum Likelihood method can be used in order to reduce the $E$-to-$B$ leakage. Additionally, the pseudo-$\mathcal{C}_{\ell}$ formalism can be used for high multipoles.

\begin{figure}[t!]
    \centering
    \includegraphics[scale = 0.35]{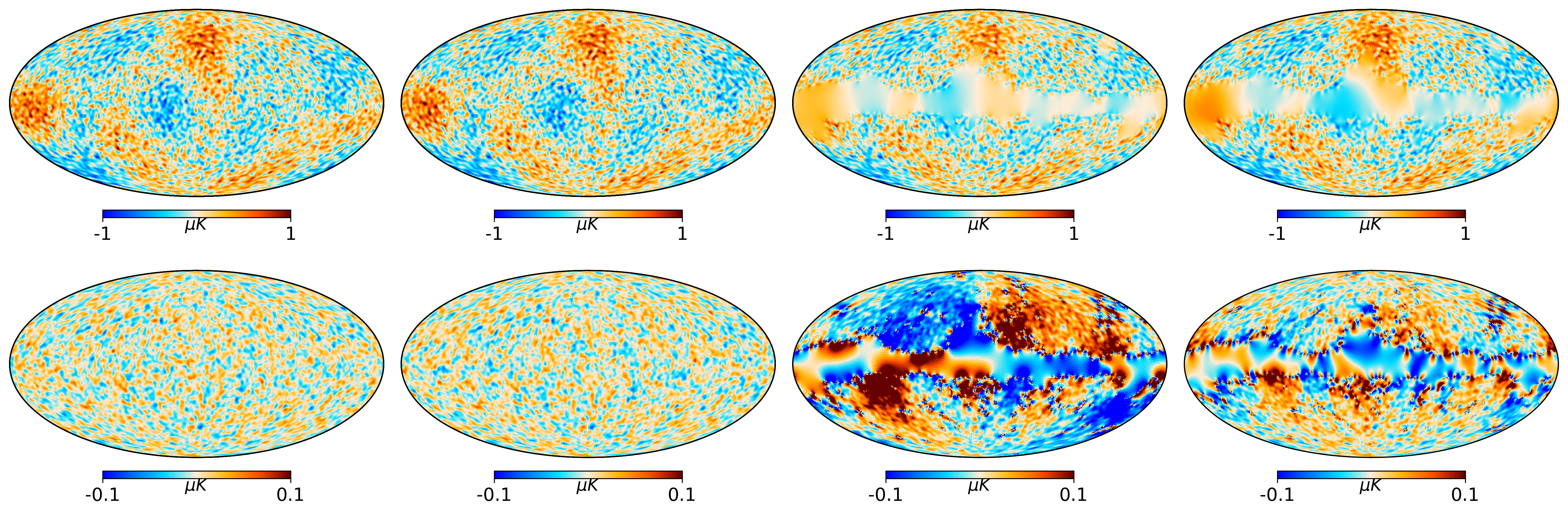}
    \caption{$E$ (first row) and $B$-mode (second row) reconstruction for different approaches. First column shows the $E$- and $B$-mode maps obtained directly from the full-sky $Q$ and $U$ maps. The recovery obtained after applying our inpainting technique in the masked area is given in the second column. The case in which pixels inside the mask are simply replaced by zeros is shown in the third column. Finally, in the last column, a diffuse inpainting approach is applied on the input $Q$ and $U$ maps before obtaining the $E$- and $B$-mode maps.}
    \label{EB_Example}
\end{figure}

As an illustration, figure \ref{EB_Example} shows the E- and B- modes maps obtained directly from Q and U full-sky maps versus those recovered after inpainting the masked region of the $Q$ and $U$ maps. For comparison, we also include the $E$- and $B$-mode maps generated directly from the masked $Q$ and $U$ maps, and those obtained after applying a diffuse inpainting technique\footnote{Diffuse inpainting consists on filling iteratively the masked pixels taking the average value of the neighbour pixels until convergence is reached.} on the Stokes $Q$ and $U$ parameters. In these two last cases, a strong $E$-to-$B$ leakage can be clearly appreciated.

\begin{figure}[t!]
    \centering
    \includegraphics[scale = 0.4]{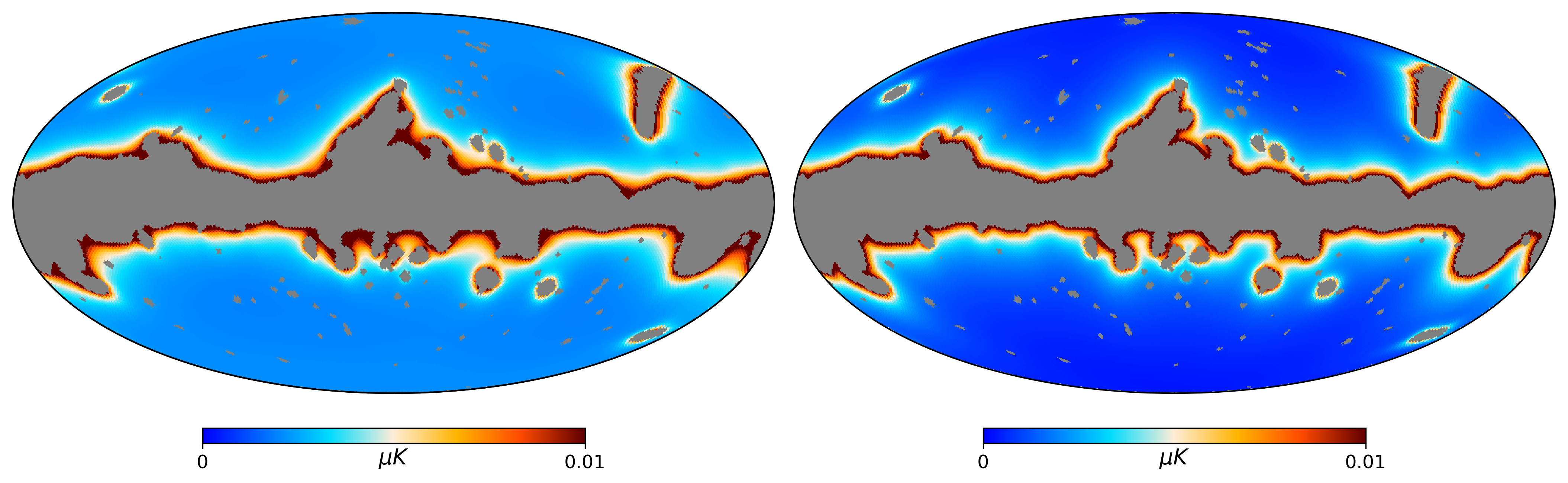}
    \caption{Standard deviation of the $E$- (left) and $B$-mode (right) residuals outside the Planck 2018 polarization confidence mask at $N_{\mathrm{side}}$ = 64.}
    \label{Std_Res}
\end{figure}

\begin{figure}[t!]
    \centering
    \includegraphics[scale = 0.5]{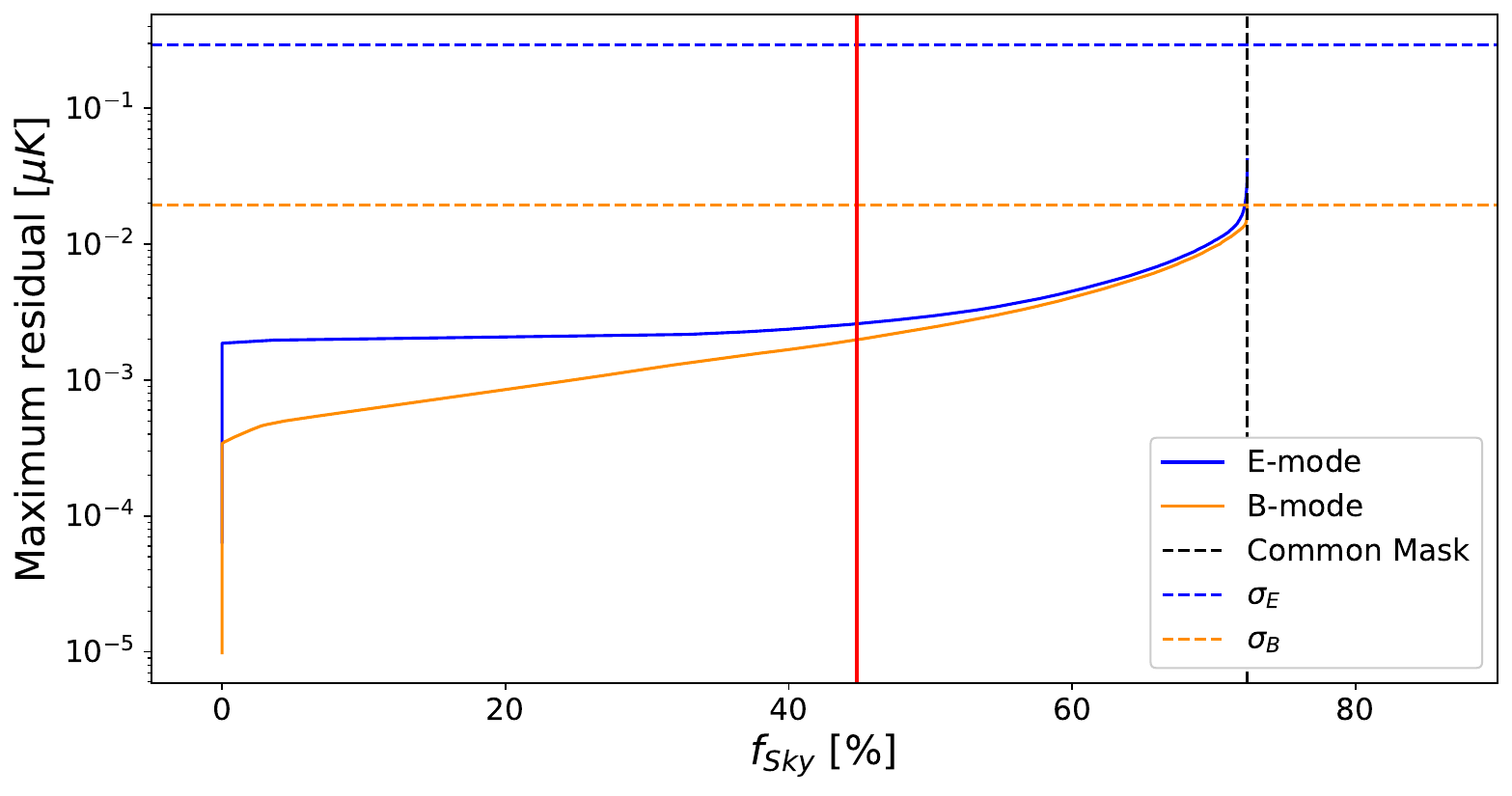}
    \caption{Evolution of the maximum residual in the $E$- and $B$-mode with respect to the available fraction of sky. The vertical black dash line corresponds to the Planck 2018 polarization common mask. The horizontal blue and orange dash lines show the typical value of the $E$- and $B$-mode fluctuations, respectively. Finally, the red solid line is fixed to $f_{sky}$ = 44.8\%, where the relative error on the B-mode reconstruction is below 10\%.}
    \label{Error_vs_Sky}
\end{figure}

In order to assess the error in the $E$- and $B$-mode reconstructions, we compute the map of the standard deviation of the residuals. Starting from the 1200 $T$, $Q$, and $U$ inpainted realizations, we generate the corresponding $E$- and $B$-mode maps and compute the residuals by subtracting from them the input $E$- and $B$-mode. We calculate then the standard deviation maps, pixel by pixel, which are shown in Figure \ref{Std_Res}. On the one hand, for the $E$-mode, the maximum error outside the polarization common mask is 0.042 $\mu{K}$ which corresponds to around a 14 per cent of the typical amplitude of the E-mode signal, $\sigma_{E}\sim 0.29$ $\mu{K}$ at the considered resolution.
On the other hand, for the $B$-mode, the maximum error is at the level of 0.019 $\mu{K}$, which is approximately the expected amplitude of the B-mode signal for r=0, $\sigma_{B}$. However, as figure \ref{Error_vs_Sky} shows, the error decreases rapidly, and for 60\% of the sky the maximum error is at the level of 20 per cent relative to $\sigma_{B}$. Concerning the $E$-mode, it is interesting to point out that the error is below 5\% for more than the 71.7\% of the sky, and then it goes down until it reaches a plateau, where we can not push the maximum residual to a lower value even if we extend the mask.

We get similar results for the absolute error in the $E$- and $B$-mode reconstruction when we include the semi-realistic noise realization. 

\subsection{Power Spectra}

\begin{figure}[t!]
    \centering
    \includegraphics[scale = 0.3]{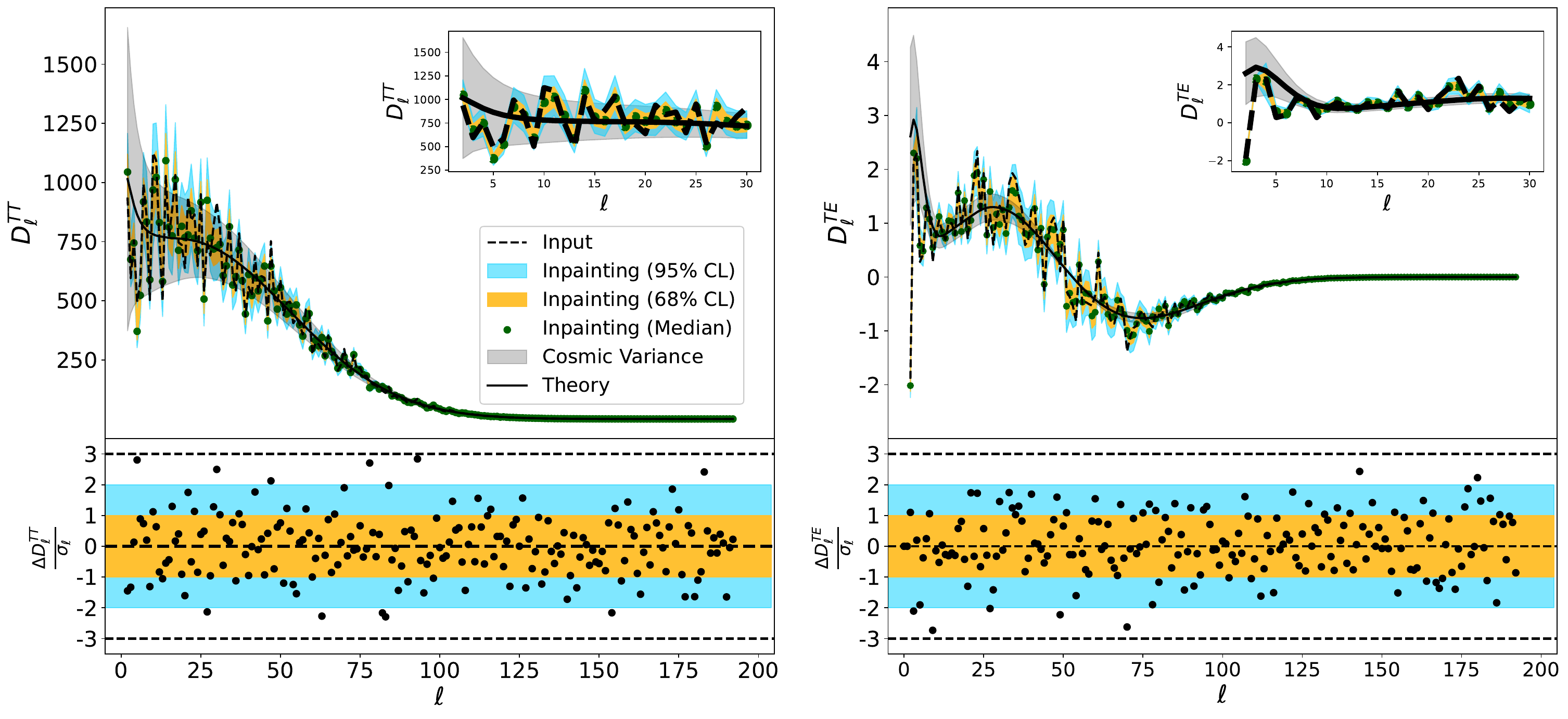}
    \caption{$TT$ (left) and $TE$ (right) power spectrum scaled by $\ell(\ell+1)/(2\pi)$ ($D_{\ell}$). The solid black line shows the input theoretical model, while the grey area corresponds to the cosmic variance. The dashed black line shows the power spectrum from the input noiseless CMB realization. Green dots correspond to the average value from the 1200 inpainted realizations, while orange and blue contours are the 68\% and 95\% C.L., respectively. Residuals are shown in the lower panel.}
    \label{TT_TE}
\end{figure}

\begin{figure}[t!]
    \centering
    \includegraphics[scale = 0.3]{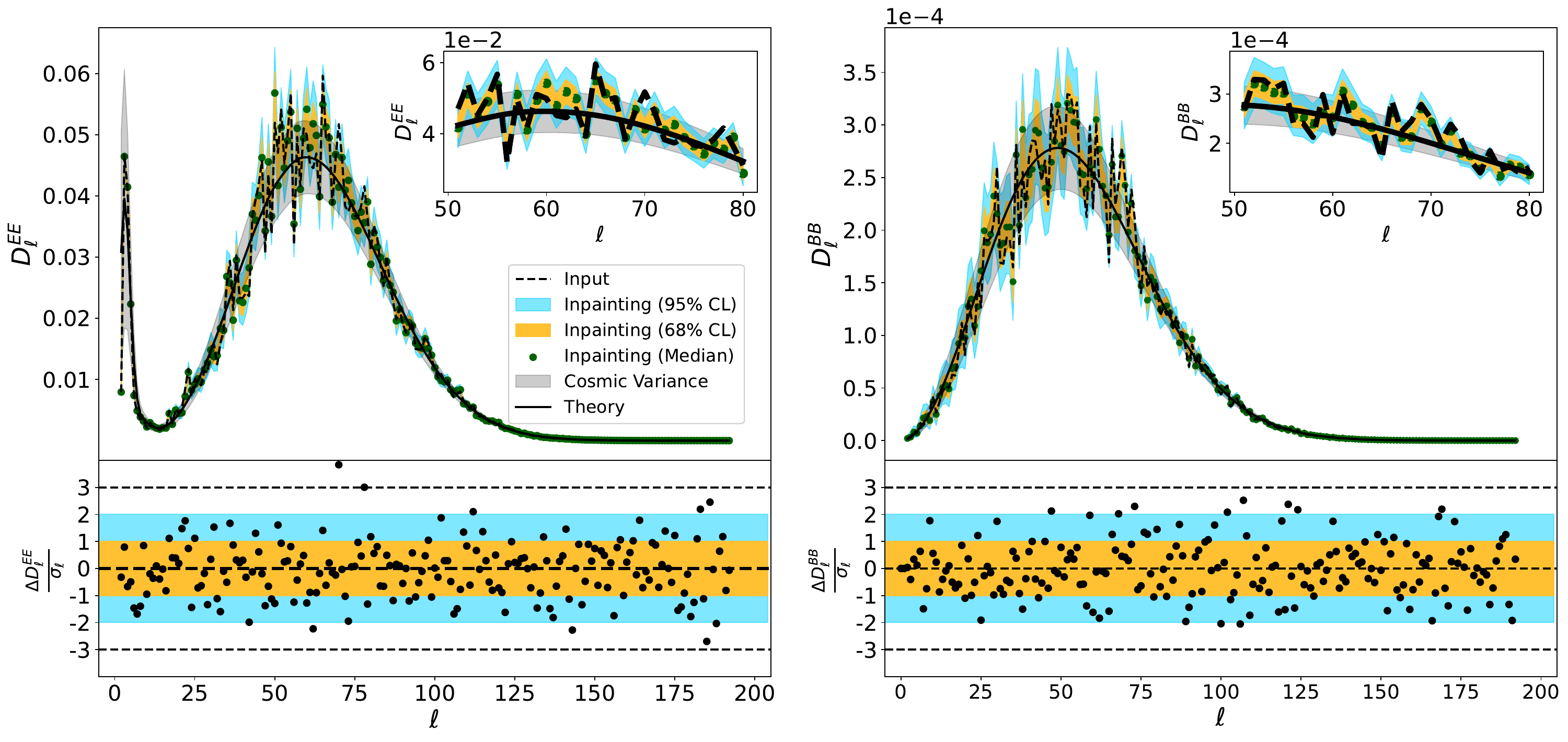}
    \caption{Same as figure \ref{TT_TE}, but for the $EE$ (left) and $BB$ (right) power spectrum.}
    \label{EE_BB}
\end{figure}

Our final test is related to the power spectra estimation. Given the input $T$, $Q$, and $U$ full-sky maps, we calculate the $TT$, $EE$, $BB$ and $TE$ power spectrum, and compare it to the mean power spectra generated from the 1200 inpainted realizations. In particular, for this case where 30 per cent of the sky is inpainted, we find that the distribution of the values for each multipole of the inpainted realizations closely resembles a Gaussian distribution, even for $\ell < 30$. For larger $f_{sky}$ to be inpainted, we expect the distribution to become more like a $\chi^2$. We also calculate the 68\% and 95\% C.L., as well as the residuals per multipole. The residuals are computed by taking the differences between input and median values over the 1200 realizations. Then, we divide by the upper or lower sigma\footnote{Low sigma is computed by integrating from the median to the 16\% of the low tail, i.e. it encapsulate 34\% of the probability. The other 34\% is in the upper sigma which is the integration between the median up to the 84\%. For instance, if the residuals are positive, which means that the input value is below the median, we use the low sigma.} to take into account possible asymmetries in the distributions, especially for low multipoles. Figure \ref{TT_TE} shows the $TT$ and $TE$ power spectra, while in figure \ref{EE_BB} $EE$ and $BB$ are plotted. Taking into account that we have a single CMB realization, the input spectra is noisy compared to the theoretical prediction due to the cosmic variance, both of them also plotted. Indeed, the mean power spectra obtained from the inpainted simulations follow that of the input CMB, rather than that of the theoretical model, finding that almost all recovered multipoles fall within the 95\% C.L. Note that for the case of the B-mode, we are considering a scenario with r=0 (corresponding to the PR3 $\Lambda$CDM best fit model) and, therefore, only the contribution from lensing is present. However, we have checked that the power spectra is equally well recovered when starting with a simulation with $r$ different from zero. 

\begin{figure}[t!]
    \centering
    \includegraphics[scale = 0.3]{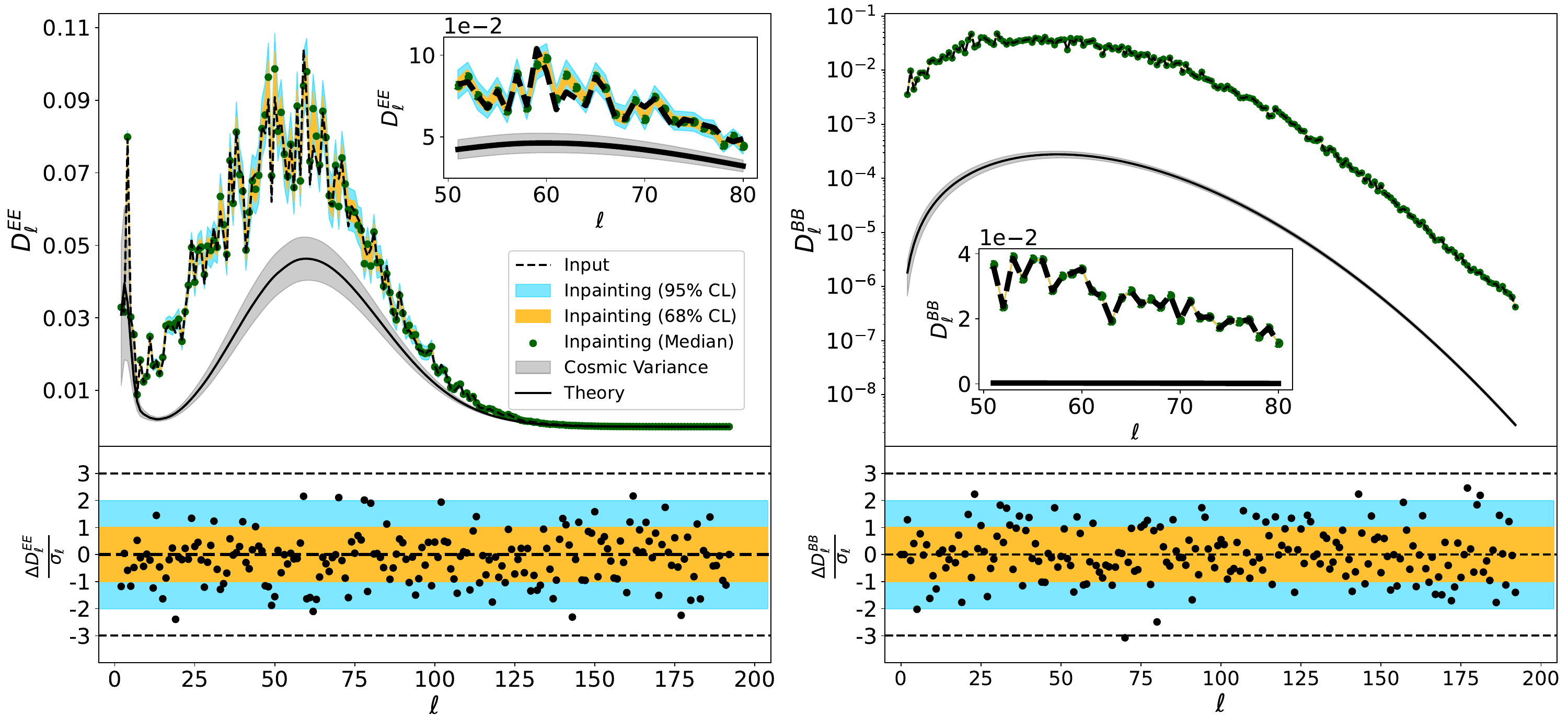}
    \caption{Same as figure \ref{EE_BB}, but for the scenario in which CMB plus semi-realistic noise is considered. As reference, we also plot the power spectrum and the cosmic variance for the CMB signal.}
    \label{EE_BB_Noise}
\end{figure}

Regarding the scenario where semi-realistic noise is included, we also see a good agreement between the input $TT$, $TE$, $EE$, and $BB$ power spectra and the ones recovered from the inpainted realizations. Note that, as one would expect, in this case the recovered power spectra for EE and BB is above that of the polarization CMB signal (see figure \ref{EE_BB_Noise}), due to the fact that the noise is the dominant contribution of the maps.

Finally, we perform a last test at the $B$-mode power spectrum level to compare between our inpainting, diffuse inpainting and a simple masking approach (i.e. put to zero all pixels inside the mask in the $Q$ and $U$ maps and then transform to $E$ and $B$). 
We start by generating $B$-mode maps from the different approaches (GCR, diffuse inpainting and masking) and mask them with the Planck polarization common mask. We compute then the corresponding power spectrum using the \texttt{PyMaster} package, the Python implementation of the \texttt{NaMaster} \cite{Alonso:2018jzx} library, which computes the angular power spectrum of a masked field using the pseudo-$\mathcal{C}_{\ell}$ formalism. In particular, the pseudo-$\mathcal{C}_{\ell}$ are computed using a C2 apodization with 15 degrees and a uniform binning including 4 multipoles per bin. 

\begin{figure}[t!]
    \centering
    \includegraphics[scale = 0.7]{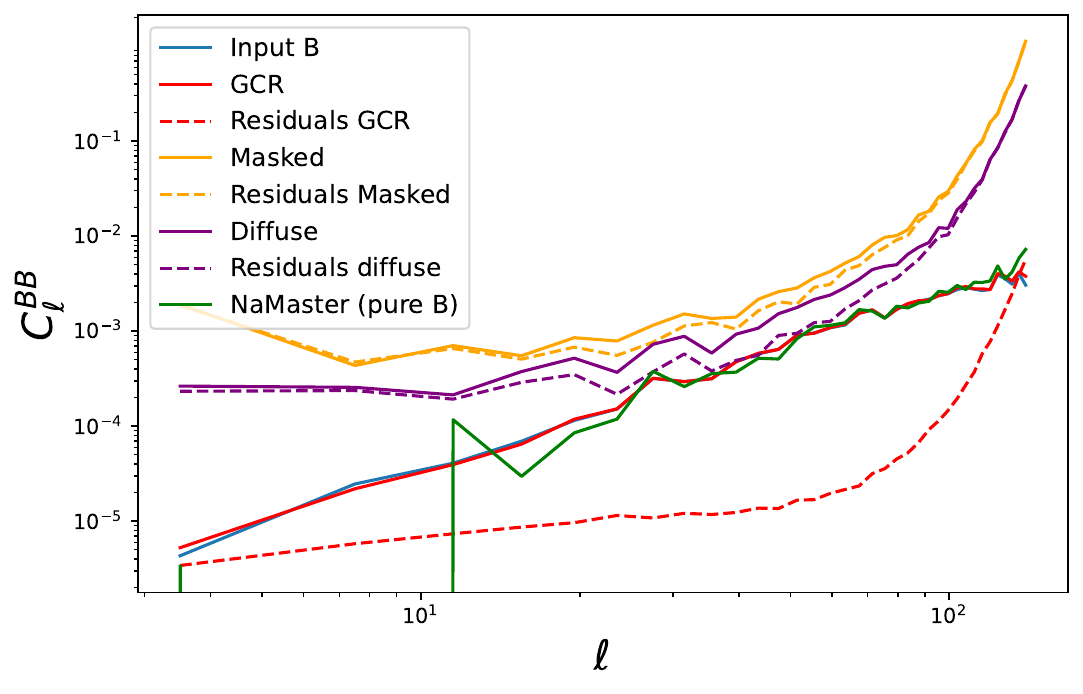}
    \caption{Pseudo-$\mathcal{C}_{\ell}$ for the $B$-mode maps reconstructed from different approaches. Residuals are computed using \texttt{NaMaster} and taking as input the difference map. Planck 2018 polarization confidence mask is used. For comparison, the solid green line shows the pseudo-$\mathcal{C}_{\ell}$ computed from masked $Q$ and $U$ maps by applying the \texttt{NaMaster} pure-$B$ technique.}
    \label{NaMaster}
\end{figure}

Figure \ref{NaMaster} shows the results. As expected, for a deep transition in the edge of the mask, i.e. the case where pixels inside the mask are replaced by zeros, the $B$-mode power spectrum is completely dominated by the $E$-to-$B$ leakage (dark orange curve). This leakage can be reduced by one order of magnitude applying a diffuse inpainting as it smooths the discontinuity in the $Q$ and $U$ maps. However, the lensing signal is still hidden below the leakage. Our results show that the GCR is the best approach to recover the input power spectrum having residuals below the signal for all the multipole range. For comparison, we also consider the \texttt{NaMaster} pure $B$ approach, which recovers the B-mode power spectrum starting from the masked $Q$ and $U$ maps. For the considered case, this approach also fails reproducing the large angular scales of the B-mode. These results show that obtaining the power spectra from an inpainted map could be used as an alternative to more standard methods. However, further work is needed to validate the usefulness of this approach.

\section{CMB-PAInT} \label{PythonCode}

As part of this work, we have developed a Python package called \texttt{CMB-PAInT} (\textbf{Cos}mic \textbf{M}icrowave \textbf{B}ackground \textbf{P}olarization \textbf{A}nisotropies \textbf{In}painting \textbf{T}ool) to perform inpainting on an input map in the HEALPix format \cite{Gorski:2004by}. We will make this user-friendly package publicly available in \url{https://github.com/ChristianGim/CMB-PAInT}.

In this section we briefly describe the software capabilities.
As an example, we also give computational times for the configuration we use at NERSC\footnote{National Energy Research Scientific Computing Center (NERSC), \url{https://www.nersc.gov/}, is a primary scientific computing facility operated by the Lawrence Berkeley National Laboratory, located in California. It provides high-performance computing and storage facilities where Planck latest data and simulations can be found.} to perform the inpainting applications shown in Section \ref{Test_Validation}. In particular, we consider the case where the inpainting is performed on the Planck temperature and polarization common masks region (14113 and 13583 pixels to be inpainted, respectively) on $T$, $Q$, and $U$ components at $N_{\mathrm{nside}}$=64.

\texttt{CMB-PAInT} can be used in different ways:

\begin{itemize}
    \item To compute the pixel covariance matrix from an input angular power spectrum up to a certain $\ell_{max}$. Depending on the field to be inpainted, it can compute the covariance matrix of either $T$, $QU$, or $TQU$. 
    \item To compute the Cholesky decomposition from an input covariance matrix. The matrix can be just signal, previously computed from an input power spectrum, or the sum of signal plus some extra component (noise, systematics...). Given the mask it also performs the required permutations in rows and columns, i.e. it orders first the unmasked pixels and then the masked ones as explained in Section \ref{GCR}.
    \item To compute the 
    \textcolor{blue}{$z$} variable and inpaint the map. If the input is a single sky map, it can generate N different inpainted realizations of the same sky. If the input is a set of maps, it computes for each of them the 
    $z$ variable and an inpainted realization. It also includes an optional parameter, \texttt{Cons\_Uncons}, to allow one saving the constrained and unconstrained parts of the inpainting process. If True, they are included in the 0 and 1 fields of the output fits file, while field 2 contains the sum.
\end{itemize}

This code can run on a NERSC-like cluster that uses \texttt{slurm} scheduling, or on a local machine, a Jupyter Notebook, or another cluster with different resource management. The only requirement is to consider the memory limitation. The code needs a configuration file that contains all the model and software parameters for running it. For example, the configuration file specifies whether to inpaint the polarization field or not, or whether to use an external .sh file. The code creates an instance of the CMB-PAInT class and runs one of the methods based on the user input: \texttt{calculate\_covariance}, \texttt{calculate\_cholesky}, or \texttt{calculate\_inpainting}. 
If the methods run on a cluster without an input .sh file, it generates a .sh file based on NERSC with the resources from the configuration file, such as number of nodes, tasks, CPUs per task, time limits, email address, or partition. The code has two levels of parallelization. It distributes the work among a number of jobs ($N_{\mathrm{jobs}}$) that are submitted together, and each job uses \texttt{mpi4py}, the MPI standard for Python, to parallelize the assigned rows or maps. This parallelization is used to calculate the covariance matrix from an input power spectrum and also to inpaint the maps. In the latter, the total number of realizations are distributed among the jobs and the tasks per job. Additionally, the code uses the \texttt{dask} package to perform Cholesky decomposition faster than the standard \texttt{numpy} implementation. The code saves intermediate products in a numpy file format, and Cholesky decomposition in a HDF5 binary data format.

\begin{figure}[t!]
    \centering
    \includegraphics[scale = 0.7]{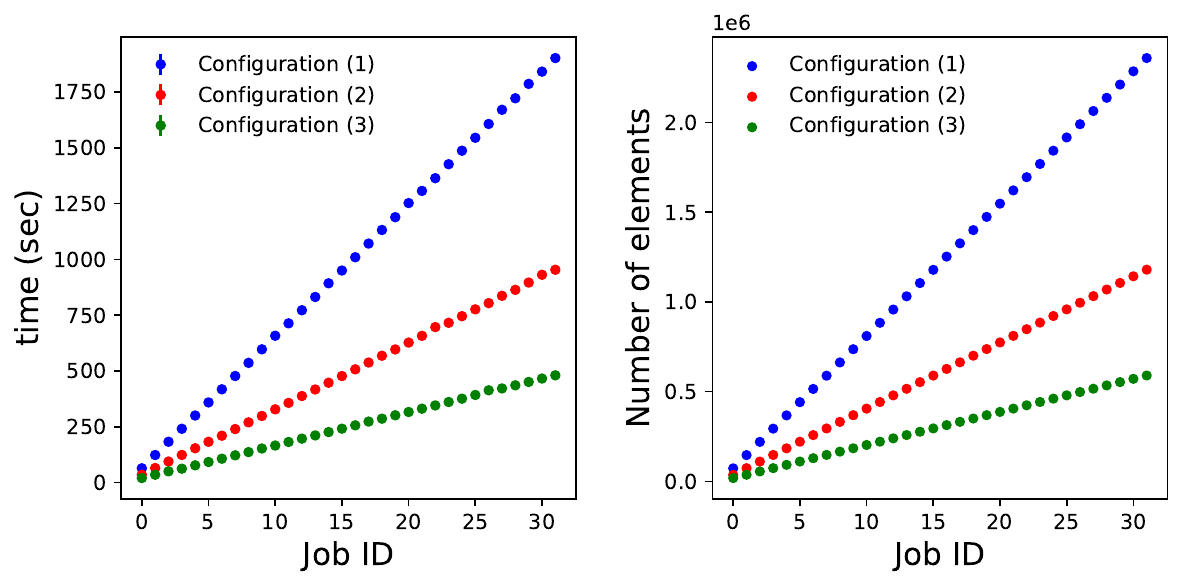}
    \caption{\textit{Left panel:} Computational time to calculate the covariance matrix for 3 different setups. \textit{Right panel:} Number of elements of the last task of each job for the three configurations.}
    \label{Computing_time}
\end{figure}

Regarding computational time, the left panel of Figure \ref{Computing_time} shows the time taken to compute the covariance matrix for a map of $N_{\mathrm{side}}$ = 64. We use the following configurations: 

\begin{enumerate}
    \item $N_{\mathrm{jobs}}$ = 32 (Single node, 32 tasks, 4 CPUs per task)
    \item $N_{\mathrm{jobs}}$ = 32 (Single node, 64 tasks, 2 CPUs per task)
    \item $N_{\mathrm{jobs}}$ = 32 (Single node, 128 tasks, 1 CPUs per task)
\end{enumerate}

Since the covariance matrix is symmetric, we only need to compute the first (i) elements of the $i^{th}$ row, which are the subdiagonal and diagonal elements. As expected, time cost increases linearly with the Job ID, because the number of operations increases in the same way. For the proposed configurations, each job computes 1536 rows ($N_{\mathrm{pix}}/N_{\mathrm{jobs}}$), which are split among the number of tasks. For instance, in configuration (1), the first task of the first job computes the rows between 0 and 47 (1176 elements), while the last task computes the rows between 1488-1535 (72600 elements). It is straightforward to conclude that the bottleneck of each job is the last task, which computes the largest number of operations. In the right panel of figure \ref{Computing_time} we display the number of elements/operations that are done by the last task of each job. For the configuration (2) the time cost is reduced by almost a factor of 2, as we assign more tasks per job and leave fewer elements to the last task. In this sense, the optimal configuration is (3). However, giving some CPUs per task could be necessary due to memory issues. Additionally, the code could have an extra paralelization layer for configurations (1) and (2) if OpenMP API is used, which will improve the performance. 

\begin{figure}[t!]
    \centering
    \includegraphics[scale = 0.8]{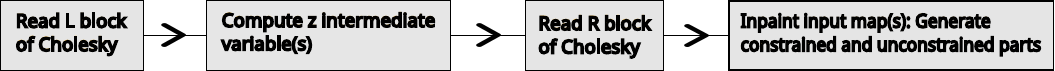}
    \caption{Diagram of the workflow in \texttt{CMB-PAInT} for the inpainting algorithm.}
    \label{Diagram}
\end{figure}

Figure \ref{Diagram} shows a diagram of the workflow for the inpainting method once the covariance matrix and the Cholesky decomposition\footnote{It takes around 30 to 40 minutes to compute the covariance matrix and the Cholesky decomposition in a Perlmutter node for the considered example:
$N_{\mathrm{side}}$=64, all the components ($T$, $Q$, and $U$) and Planck common masks.} are computed. In this case, we try the following configuration:

\begin{itemize}
    \item  $N_{jobs} = 16$ (3 nodes per job, 10 tasks, 4 CPUs per task)
\end{itemize}

\begin{figure}[t!]
    \centering
    \includegraphics[scale = 0.45]{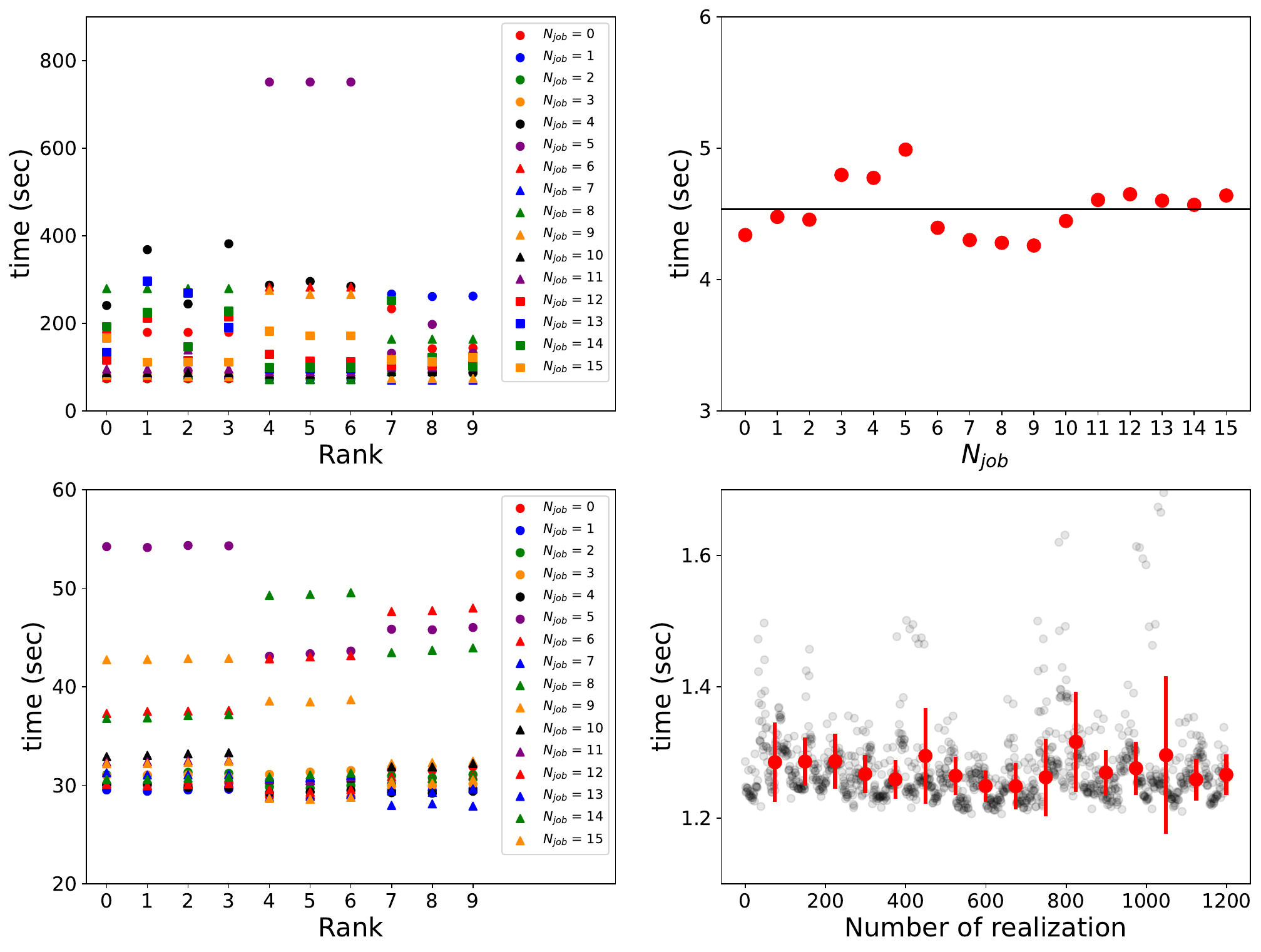}
    \caption{Computational time of each of the steps in Figure \ref{Diagram}. \textit{Upper left:} Computational time per rank and per job to read the L matrix. \textit{Upper right:} Computational time per job to compute the z variable. \textit{Lower left:} Computational time per rank and per job to read the R matrix. \textit{Lower right:} Computational time to inpaint each realization (open grey dots). In red the mean time and dispersion per job (note that each job inpaints $N_{\mathrm{sims}}$/$N_{\mathrm{jobs}}$ realizations).}
\label{Inpainting_time}
\end{figure}

Figure \ref{Inpainting_time} shows the time cost of each step in the workflow. The bottleneck is the reading of the L matrix, which is the largest block in the Cholesky decomposition. It takes less than 300 seconds for most ranks\footnote{In the MPI context, every process that belongs to a communicator is uniquely identified by its rank, which is an integer that ranges from zero up to the size of the communicator minus one.} and jobs. However, some ranks get stuck and take more than 700 seconds to read it. The next step is computing recursively the $z$ variable, which takes around 4.5 seconds with slight variations. Reading the R matrix is faster than reading the L matrix, but there is a large variation in this run. On average, this step takes around 30 seconds. The last step is inpainting a map, which takes less than 1.3 seconds on average.

\section{Conclusions} \label{Conclusions}

In this work we have presented an inpainting technique based on Gaussian Constrained Realizations, that can be applied to CMB temperature and polarization data. The algorithm uses the Cholesky decomposition to sample from a conditional probability distribution. We have also developed a \texttt{Python} user-friendly package, \texttt{CMB-PAInT}, which will be publicly available in \url{https://github.com/ChristianGim/CMB-PAInT}. We used this package to obtain all the inpainted realizations for this paper.

In order to asses the performance of the methodology, a series of tests have been done in two different scenarios: (1) CMB signal only, and (2) CMB signal with semi-realistic noise simulations based on Planck Release 4.
In particular, we checked that the constrained part of the inpainted maps was consistent with that expected from the model. We also studied the one-dimensional probability distribution of $T$, $Q$, and $U$, as well as those of the $E$- and $B$-modes. In both scenarios, they agreed well with the input values within the expected errors. Our methodology is also able to reconstruct accurately systematics and noise from an anisotropic field if they are included in the covariance matrix. However, this requires that all the correlations between pixels of the anisotropic field are perfectly characterised. Otherwise, artifacts would appear due to a mismatch between the statistical properties of the pixels outside the mask and the assumed model. This can be checked by inspecting the intermediate $z$ variables, which should follow a normal distribution if everything is consistent. 

For certain applications, further studies may be needed in order to minimise the presence of prominent systematics in the inpainted region while, at the same time, not introducing a statistical mistmatch between observed pixels and the covariance matrix. In any case, the optimal strategy will depend on the nature of the data and estimator to be studied.

The most interesting tests involve the $E$- and $B$-mode reconstruction and the corresponding power spectra. Both tests show that we can remove well the $E$-to-$B$ leakage. For the first scenario, at the map level, we are able to reconstruct the $E$-mode map with a relative error below 5\% for a sky fraction of 71.7 per cent, covering almost all the sky outside the polarization confidence mask (which allows 72.36 per cent of the pixels). In the case of the $B$-mode, the relative error is around 10\% for $f_{sky}$ = 45\%, due to its weaker signal. At the power spectra level, we reproduce the input $TT$, $TE$, $EE$, and $BB$ power spectra up to $\ell_{max}$ = 192, covering the full range of the reionization and the recombination peaks of the $B$-mode. The residuals between the input and reconstructed spectra are 
consistent with those expected from the dispersion obtained from the 1200 inpainted realizations. In the second scenario, the method reproduces the input power spectra that were strongly affected by systematics and noise. Furthermore, we perform a comparison between our methodology and other techniques (diffuse inpainting, masking and the NaMaster pure-B approach), finding that the GCR was the only method able to recover the input $B$-mode spectrum for all the considered multipole range. 

This inpainting approach is limited to low resolution maps due to computational memory requirements, but it is enough to target the polarization largest scales, which is the main goal of future observations, searching for the primordial B-mode. In view of the present results, we believe that this will be a useful and powerful algorithm for the analyses of future CMB experiments, such as LiteBIRD \cite{LiteBIRD:2022cnt}.

\appendix
\section{Convergence of the covariance matrix} \label{Appendix_A}

\begin{figure}[t!]
    \centering
    \includegraphics[scale = 0.44]{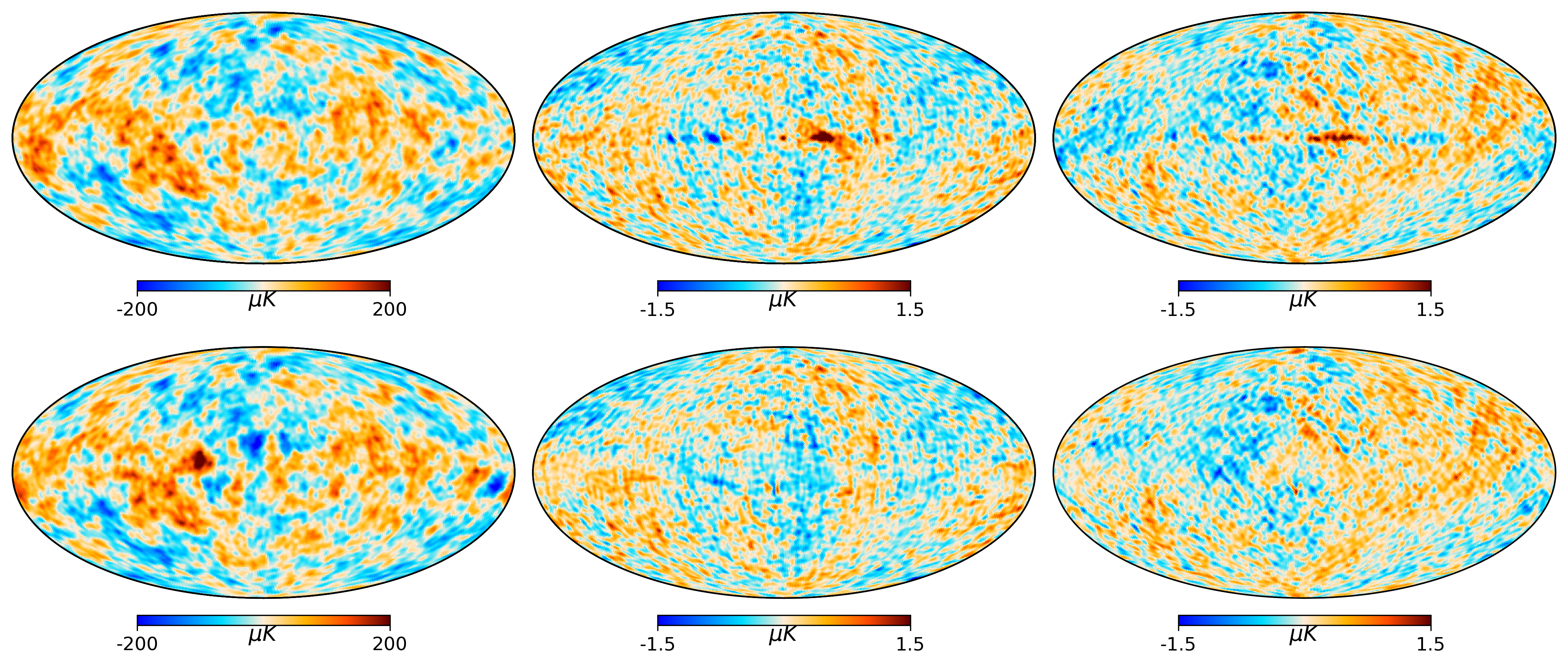}
    \caption{Example of a $T$ (left), $Q$ (middle), and $U$ (right) inpainted realization in the case where 20 semi-realistic noise simulations are used to characterize the noise covariance matrix. The top and bottom panels correspond to the input and inpainted maps, respectively.}
    \label{Example_20}
\end{figure}

In section \ref{Test_Validation} we show the performance of the inpainting for two different scenarios: (1) CMB signal only and (2) CMB signal plus a semi-realistic noise realization. In the second case, the methodology is able to reconstruct accurately the systematics and noise. However, this is only possible if the full anisotropic covariance matrix is well characterized. In a more realistic case, where a limited number of simulations are used to compute the matrix numerically, the non-convergence of the matrix induces a mismatch between the matrix and the pixels outside the mask. This introduces some artifacts in the inpainted realizations. Figure \ref{Example_20} shows an example of the input and output $T$, $Q$, and $U$ maps using just 20 semi-realistic noise simulations to compute the covariance matrices. Even if the mismatch is mainly for the polarization field\footnote{For temperature the noise is much smaller than the signal and, even if it were not well modelled in the covariance matrix, its effect is negligible as the error in the $z$ variables is masked by the regularization noise.}, strong cold and hot spots are induced in the inpainted temperature map  through the TE correlation. 

\begin{figure}[t!]
    \centering
    \includegraphics[scale = 0.4]{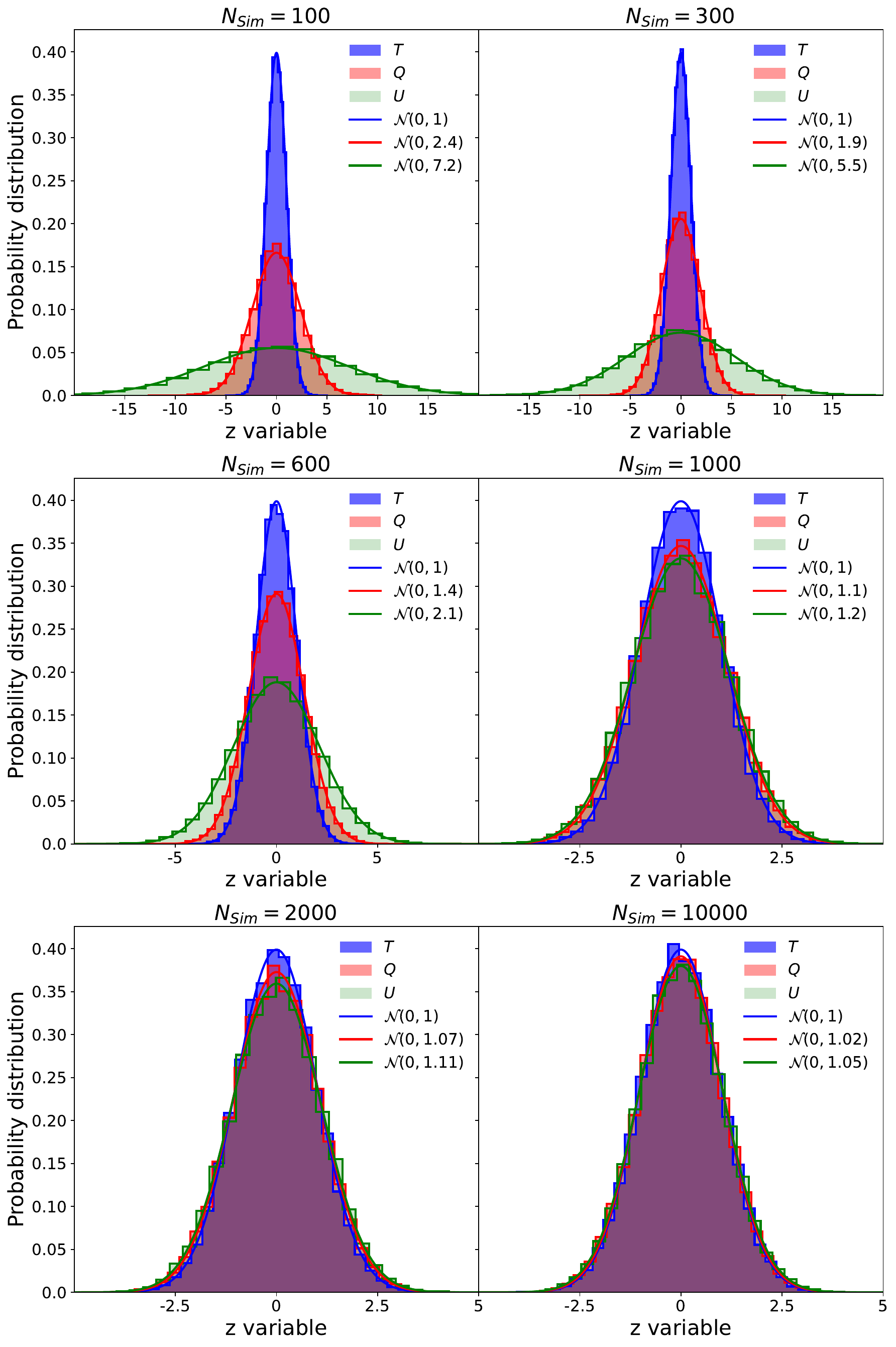}
    \caption{Distribution of the $z$ variable associated to the $T$, $Q$, and $U$ fields for different number of simulations used to estimate the noise covariance matrix. For comparison, a Gaussian with the same standard deviation is also given in each case.}
    \label{Z_distribution}
\end{figure}

The mismatch is even more clear in Figure \ref{Z_distribution}, where the probability distribution of the $z$ variables (see eq. \ref{Eq:2.12}) is plotted for $T$, $Q$, and $U$. For comparison, a Gaussian curve with the same standard deviation is plotted. In all the cases shown in figure \ref{Z_distribution}, the variable $z_{T}$, corresponding to the temperature field, follows a Gaussian distribution with zero mean and unit variance, $\mathcal{N}$(0, 1). This is because there is not a mismatch between the statistical properties of the pixels and those encoded in the matrix. In this case, even if we are not including correctly the noise and systematics on the matrix, the mismatch is simply masked by the regularization noise. However, the mismatch strongly affects the polarization field. In particular, since $Q$ and then $U$ are computed recursively, this effect is most notable for the $z$ variables associated to $U$, which is found to be the broadest one. The $z_{Q}$ and $z_{U}$ distributions become effectively $\mathcal{N}$(0, 1) when several thousand of noise simulations are considered to construct the matrix.

\section{Robustness against the model} \label{Appendix_B}

In section \ref{Test_Validation} we generate a pixel covariance matrix that match perfectly the input simulation. In a real situation this is not possible. Here, we introduce a small mismatch between the model used for the input simulation and the one used for the pixel covariance matrix estimation. As described in \ref{GCR}, we use the best fit to the $\Lambda$CDM model to generate the input simulation (only CMB is included in the simulation). We introduce a small deviation in the input parameters taking into account the correlation matrix of the estimated parameters inferred from the Planck 2018 data. In Table \ref{Params} we show the best fit parameters and the modified ones. The alternative parameters were obtained as a Gaussian random realisation of the parameters centred in the best-fit model and following the correlation matrix. Therefore, they are also consistent with the Planck data within the estimated errors.

\begin{table}[t!]
\centering
\begin{tabular}{c|c|c|}
\cline{2-3}
                                               & $\Lambda CDM$ best fit & Modified $\Lambda CDM$ \\ \hline
\multicolumn{1}{|c|}{$\Omega_{b}h^{2}$}        & 0.02238280             & 0.02244959             \\ \hline
\multicolumn{1}{|c|}{$\Omega_{c}h^{2}$}        & 0.1201075              & 0.1194516              \\ \hline
\multicolumn{1}{|c|}{$H_{0}$}                  & 67.32117               & 67.57953               \\ \hline
\multicolumn{1}{|c|}{$\tau$}                   & 0.05430842             & 0.05744905             \\ \hline
\multicolumn{1}{|c|}{${\rm{ln}}(10^{10} A_s)$} & 3.044784               & 3.056551               \\ \hline
\multicolumn{1}{|c|}{$n_{s}$}                  & 0.9660499              & 0.9654136              \\ \hline
\end{tabular}
\caption{Cosmological parameters. $\textit{Left}$: $\Lambda CDM$ best fit. Model used for the input simulation. $\textit{Right}$: Modified model according to the Planck 2018 errorbars and correlations.}
\label{Params}
\end{table}

\begin{figure}[t!]
    \centering
    \includegraphics[scale = 0.44]{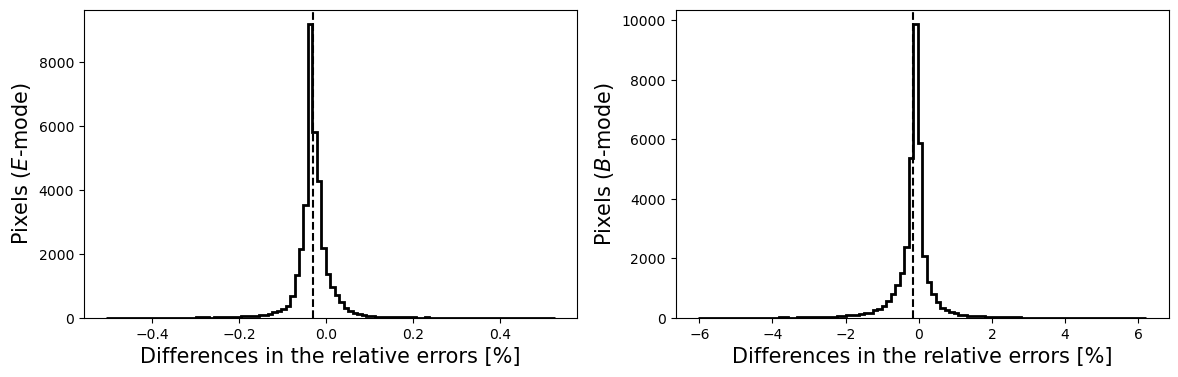}
    \caption{Distribution of differences of the errors relative to the cosmic variance of the exact minus the modified model for $E$-mode ($\textit{left}$) and $B$-mode ($\textit{right}$).}
    \label{Differences_Errors}
\end{figure}

We compute the relative error with respect cosmic variance in the $E$- and $B$-mode reconstruction using the 1200 inpainted realizations, as explained in section \ref{EB_recon}, for both cases, the exact and the modified model. Then, we get the differences between previous errors, which are on average at the level of 0.1\% (errors are slightly larger in the modified case) as it is shown in Figure \ref{Differences_Errors}. 

\begin{figure}[t!]
    \centering
    \includegraphics[scale = 0.6]{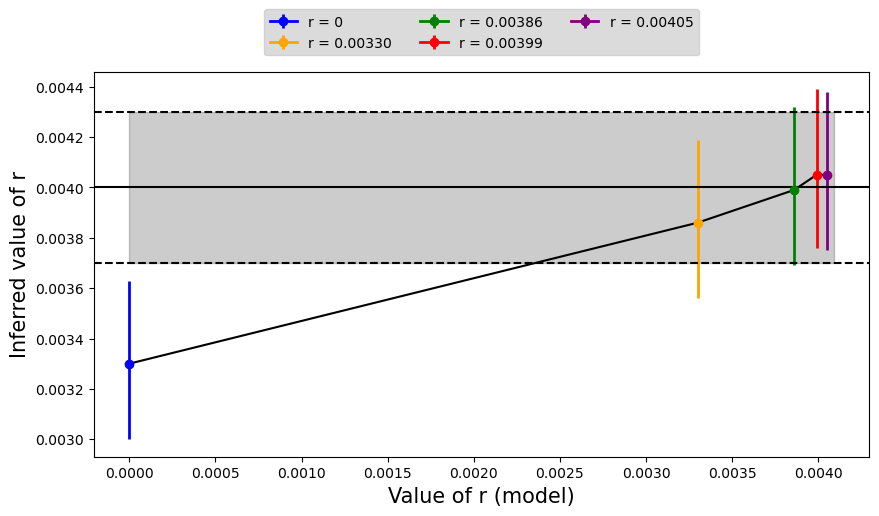}
    \caption{Value of r assumed in the fiducial model versus the inferred value. Grey shaded region corresponds to the 1$\sigma$ interval of the posterior for the input simulation, which is generated assuming a tensorial modes with r = 0.004. Solid black line corresponds to the value where the posterior peaks for the input simulation, which by chance corresponds to the same value as the model. Dots corresponds to the values of r where the mean posterior peaks for each assumed model in the iterative process. Error bars are the 1$\sigma$ interval of the mean posterior.}
    \label{r_inference}
\end{figure}
Similar results are found at the power spectra level. This is expected taking into account that the differences between models in this multipole range ($\ell$ = 2-192) is much smaller than the cosmic variance, so smaller than the differences between different realizations of the same model.

As an extra test, we consider a special scenario where tensor-to-scalar ratio (r) is equal to 0.004. In this case, most of the constraining power is encoded in the $B$-mode large scales, so differences in the model can impact and bias the r estimation. We demonstrate that an iterative process can be a good approach to this issue. Taking into account that tensorial modes have not been detected yet, it is reasonable to start with a model where r = 0. We generate 1200 inpainted realizations assuming r = 0 model
and we estimate for each of them the posterior (using a flat prior) of r using the following exact likelihood in harmonic space:

\begin{equation}
    -\log{\mathcal{L}} = \sum_{\ell}\left[\frac{\hat{\mathcal{C}}_{\ell}}{\mathcal{C}_{\ell}} + \log{\mathcal{C}_{\ell}}-\frac{2\ell-1}{2\ell+1}\log{\hat{\mathcal{C}}_{\ell}}\right]
\end{equation}
where $\hat{\mathcal{C}}_{\ell}$ is the $B$-mode power spectrum of the realization, $\mathcal{C}_{\ell}$ is the theoretical spectrum, and the sum is done up to $\ell$ = 2$N_{\mathrm{side}}$. We generate the mean posterior by averaging the -$\log{\mathcal{L}}$, and then, we use the value of r where the mean posterior peaks to generate the model for the next iteration. Results for the different iterations are shown in Figure \ref{r_inference}, where the blue contour corresponds to the 1$\sigma$ interval obtained from the posterior distribution of the input simulation (a uniform prior is used for r). Red error bars corresponds to 1$\sigma$ interval of the mean posterior. It becomes apparent that after a few iterations the correct value of r is recovered. Therefore, when using the inpainting technique, a comparison between the assumed model and the one recovered from the inpainted maps is recommended in order to test the consistency of the results. 

\begin{figure}[t!]
  \centering
  \includegraphics[width=0.48\textwidth]{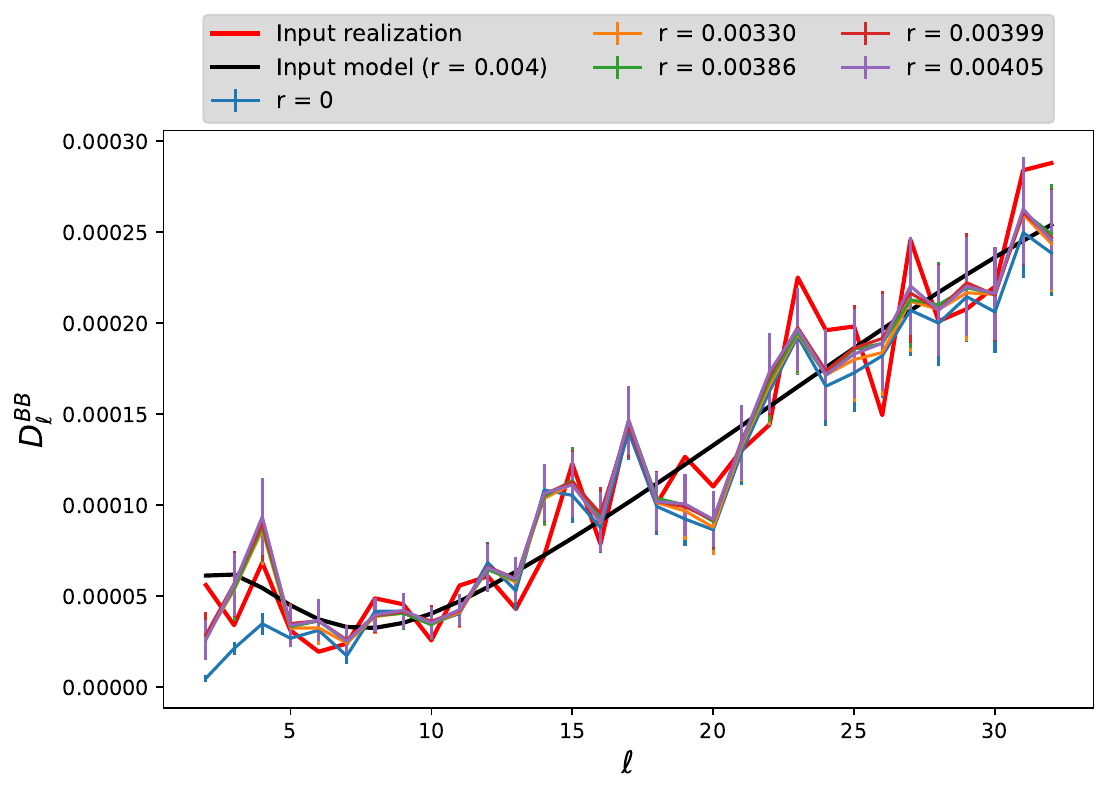}
  \includegraphics[width=0.48\textwidth]{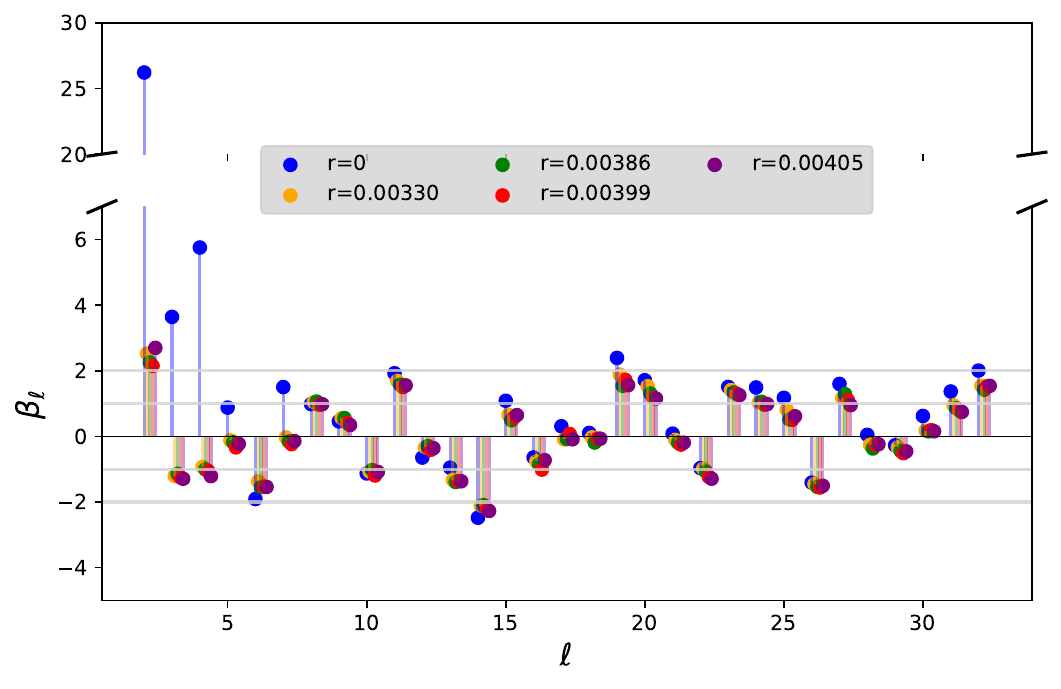}
  \caption{\textit{Left panel}: $D_{\ell}^{BB}$ for the input simulation (in red), theoretical curve for r = 0.004 (in black), and $\tilde{D}_{\ell}^{BB}$ for different fiducial models, obtained from the distribution of 1200 inpainted realizations. Error bars are also derived from the 68\% C.L of that distribution. \textit{Right panel}: $\beta_{\ell}$ for different models.}
  \label{pb_bb}
\end{figure}

We finally show a comparison between the $B$-mode power spectrum obtained using different models. Left panel of Figure \ref{pb_bb} shows $D_{\ell}^{BB}$ of the input simulation (red), the theoretical curve of the input model (black), and the estimated $\tilde{D}_{\ell}^{BB}$ for the different fiducial models (corresponding to different iterations), which are described by the median of the distribution of the 1200 spectra, obtained from each of the inpainted realizations, and the 68\% C.L (error bars). In the right panel we show $\beta_{\ell}$ defined as,
\begin{equation}
    \beta_{\ell} = \frac{x_{\ell}}{\sigma_{\ell}}
\end{equation}
where $x_{\ell}$ is the difference between the input spectrum and the median, and $\sigma_{\ell}$ is the 68\% two sided C.L, thus allowing asymmetric error bars for low $\ell$. It is clear that a bad choice of the model has a bigger impact on the first multipoles, corresponding to the reionization bump, than in the lensing dominated range.  

\acknowledgments
 
The authors would like to thank the Spanish MCIN/AEI/10.13039/501100011033, project refs. PID2019-110610RB-C21 and PID2022-139223OB-C21 (the latter funded also by European Union NextGenerationEU/PRTR) for finantial support, and the support from the Universidad de Cantabria and Consejer\'{\i}a de Universidades, Igualdad, Cultura y Deporte from the Gobierno de Cantabria through the \emph{Instrumentaci\'on y ciencia de datos para sondear la naturaleza del universo} project.
CGA also thanks the funding from the Formación de Personal Investigador (FPI, Ref. PRE2020-096429) program of the Spanish Ministerio de Ciencia, Innovaci\'on y Universidades. 
The presented results are based on observations obtained with Planck\footnote{\url{http://www.esa.int/Planck}}, an ESA
science mission with instruments and contributions directly funded
by ESA Member States, NASA, and Canada. This research uses resources of the National Energy Research Scientific Computing Center (NERSC), a U.S. Department of Energy Office of Science User Facility located at Lawrence Berkeley National Laboratory, operated under Contract No. DE-AC02-05CH11231. The results of this paper have been derived using the \texttt{HEALPix} package \cite{Gorski:2004by}, and the \texttt{healpy} \cite{Zonca2019}, \texttt{numpy} \cite{2020Natur.585..357H}, \texttt{matplotlib} \cite{2007CSE.....9...90H}, \texttt{scipy} \cite{2020NatMe..17..261V}, and \texttt{dask}\footnote{\url{https://dask.org}} \texttt{Python} packages.




\bibliographystyle{JHEP}
\bibliography{Bibliography}

\providecommand{\href}[2]{#2}\begingroup\raggedright\begin{thebibliography}{10}

\bibitem{2020A&A...641A...6P}
{Planck Collaboration}, N.~{Aghanim}, Y.~{Akrami}, M.~{Ashdown}, J.~{Aumont}, C.~{Baccigalupi} et~al., \emph{{Planck 2018 results. VI. Cosmological parameters}}, \href{https://doi.org/10.1051/0004-6361/201833910}{\emph{Astron. Astrophys.} {\bfseries 641} (2020) A6} [\href{https://arxiv.org/abs/1807.06209}{{\ttfamily 1807.06209}}].

\bibitem{2020A&A...641A..10P}
{\scshape Planck} collaboration, \emph{{Planck 2018 results. X. Constraints on inflation}}, \href{https://doi.org/10.1051/0004-6361/201833887}{\emph{Astron. Astrophys.} {\bfseries 641} (2020) A10} [\href{https://arxiv.org/abs/1807.06211}{{\ttfamily 1807.06211}}].

\bibitem{Planck:2018yye}
{\scshape Planck} collaboration, \emph{{Planck 2018 results. IV. Diffuse component separation}}, \href{https://doi.org/10.1051/0004-6361/201833881}{\emph{Astron. Astrophys.} {\bfseries 641} (2020) A4} [\href{https://arxiv.org/abs/1807.06208}{{\ttfamily 1807.06208}}].

\bibitem{2012MNRAS.420.2162F}
R.~{Fern{\'a}ndez-Cobos}, P.~{Vielva}, R.B.~{Barreiro} and E.~{Mart{\'\i}nez-Gonz{\'a}lez}, \emph{{Multiresolution internal template cleaning: an application to the Wilkinson Microwave Anisotropy Probe 7-yr polarization data}}, \href{https://doi.org/10.1111/j.1365-2966.2011.20182.x}{\emph{MNRAS} {\bfseries 420} (2012) 2162} [\href{https://arxiv.org/abs/1106.2016}{{\ttfamily 1106.2016}}].

\bibitem{2008ApJ...676...10E}
H.K.~{Eriksen}, J.B.~{Jewell}, C.~{Dickinson}, A.J.~{Banday}, K.M.~{G{\'o}rski} and C.R.~{Lawrence}, \emph{{Joint Bayesian Component Separation and CMB Power Spectrum Estimation}}, \href{https://doi.org/10.1086/525277}{\emph{Astrophys. J.} {\bfseries 676} (2008) 10} [\href{https://arxiv.org/abs/0709.1058}{{\ttfamily 0709.1058}}].

\bibitem{2008arXiv0803.1814C}
J.-F.~{Cardoso}, M.~{Martin}, J.~{Delabrouille}, M.~{Betoule} and G.~{Patanchon}, \emph{{Component separation with flexible models. Application to the separation of astrophysical emissions}}, \href{https://doi.org/10.48550/arXiv.0803.1814}{\emph{arXiv e-prints} (2008) arXiv:0803.1814} [\href{https://arxiv.org/abs/0803.1814}{{\ttfamily 0803.1814}}].

\bibitem{10.1111/j.1365-2966.2011.19770.x}
S.~Basak and J.~Delabrouille, \emph{{A needlet internal linear combination analysis of WMAP 7-year data: estimation of CMB temperature map and power spectrum}}, \href{https://doi.org/10.1111/j.1365-2966.2011.19770.x}{\emph{Monthly Notices of the Royal Astronomical Society} {\bfseries 419} (2011) 1163}.

\bibitem{2004ApJ...605...14E}
H.K.~{Eriksen}, F.K.~{Hansen}, A.J.~{Banday}, K.M.~{G{\'o}rski} and P.B.~{Lilje}, \emph{{Asymmetries in the Cosmic Microwave Background Anisotropy Field}}, \href{https://doi.org/10.1086/382267}{\emph{Astrophys. J.} {\bfseries 605} (2004) 14} [\href{https://arxiv.org/abs/astro-ph/0307507}{{\ttfamily astro-ph/0307507}}].

\bibitem{2004ApJ...609...22V}
P.~{Vielva}, E.~{Mart{\'\i}nez-Gonz{\'a}lez}, R.B.~{Barreiro}, J.L.~{Sanz} and L.~{Cay{\'o}n}, \emph{{Detection of Non-Gaussianity in the Wilkinson Microwave Anisotropy Probe First-Year Data Using Spherical Wavelets}}, \href{https://doi.org/10.1086/421007}{\emph{Astrophys. J.} {\bfseries 609} (2004) 22} [\href{https://arxiv.org/abs/astro-ph/0310273}{{\ttfamily astro-ph/0310273}}].

\bibitem{2004PhRvD..69f3516D}
A.~{de Oliveira-Costa}, M.~{Tegmark}, M.~{Zaldarriaga} and A.~{Hamilton}, \emph{{Significance of the largest scale CMB fluctuations in WMAP}}, \href{https://doi.org/10.1103/PhysRevD.69.063516}{\emph{Phys. Rev. D} {\bfseries 69} (2004) 063516} [\href{https://arxiv.org/abs/astro-ph/0307282}{{\ttfamily astro-ph/0307282}}].

\bibitem{2004MNRAS.354..641H}
F.K.~{Hansen}, A.J.~{Banday} and K.M.~{G{\'o}rski}, \emph{{Testing the cosmological principle of isotropy: local power-spectrum estimates of the WMAP data}}, \href{https://doi.org/10.1111/j.1365-2966.2004.08229.x}{\emph{MNRAS} {\bfseries 354} (2004) 641} [\href{https://arxiv.org/abs/astro-ph/0404206}{{\ttfamily astro-ph/0404206}}].

\bibitem{2006A&A...454..409B}
A.~{Bernui}, T.~{Villela}, C.A.~{Wuensche}, R.~{Leonardi} and I.~{Ferreira}, \emph{{On the cosmic microwave background large-scale angular correlations}}, \href{https://doi.org/10.1051/0004-6361:20054243}{\emph{Astron. Astrophys.} {\bfseries 454} (2006) 409} [\href{https://arxiv.org/abs/astro-ph/0601593}{{\ttfamily astro-ph/0601593}}].

\bibitem{2010AdAst2010E..77V}
P.~{Vielva}, \emph{{A Comprehensive Overview of the Cold Spot}}, \href{https://doi.org/10.1155/2010/592094}{\emph{Advances in Astronomy} {\bfseries 2010} (2010) 592094} [\href{https://arxiv.org/abs/1008.3051}{{\ttfamily 1008.3051}}].

\bibitem{2009ApJ...704.1448H}
F.K.~{Hansen}, A.J.~{Banday}, K.M.~{G{\'o}rski}, H.K.~{Eriksen} and P.B.~{Lilje}, \emph{{Power Asymmetry in Cosmic Microwave Background Fluctuations from Full Sky to Sub-Degree Scales: Is the Universe Isotropic?}}, \href{https://doi.org/10.1088/0004-637X/704/2/1448}{\emph{Astrophys. J.} {\bfseries 704} (2009) 1448} [\href{https://arxiv.org/abs/0812.3795}{{\ttfamily 0812.3795}}].

\bibitem{Planck:2019evm}
{\scshape Planck} collaboration, \emph{{Planck 2018 results. VII. Isotropy and Statistics of the CMB}}, \href{https://doi.org/10.1051/0004-6361/201935201}{\emph{Astron. Astrophys.} {\bfseries 641} (2020) A7} [\href{https://arxiv.org/abs/1906.02552}{{\ttfamily 1906.02552}}].

\bibitem{Gimeno-Amo_2023}
C.~Gimeno-Amo, R.~Barreiro, E.~Martínez-González and A.~Marcos-Caballero, \emph{Hemispherical power asymmetry in intensity and polarization for {P}lanck {PR4} data}, \href{https://doi.org/10.1088/1475-7516/2023/12/029}{\emph{Journal of Cosmology and Astroparticle Physics} {\bfseries 2023} (2023) 029}.

\bibitem{PhysRevD.64.063001}
M.~Tegmark and A.~de~Oliveira-Costa, \emph{How to measure cmb polarization power spectra without losing information}, \href{https://doi.org/10.1103/PhysRevD.64.063001}{\emph{Phys. Rev. D} {\bfseries 64} (2001) 063001}.

\bibitem{Lewis:2003an}
A.~Lewis, \emph{{Harmonic E/B decomposition for CMB polarization maps}}, \href{https://doi.org/10.1103/PhysRevD.68.083509}{\emph{Phys. Rev. D} {\bfseries 68} (2003) 083509} [\href{https://arxiv.org/abs/astro-ph/0305545}{{\ttfamily astro-ph/0305545}}].

\bibitem{Alonso:2018jzx}
{\scshape LSST Dark Energy Science} collaboration, \emph{{A unified pseudo-$C_\ell$ framework}}, \href{https://doi.org/10.1093/mnras/stz093}{\emph{Mon. Not. Roy. Astron. Soc.} {\bfseries 484} (2019) 4127} [\href{https://arxiv.org/abs/1809.09603}{{\ttfamily 1809.09603}}].

\bibitem{Tegmark:1996qt}
M.~Tegmark, \emph{{How to measure CMB power spectra without losing information}}, \href{https://doi.org/10.1103/PhysRevD.55.5895}{\emph{Phys. Rev. D} {\bfseries 55} (1997) 5895} [\href{https://arxiv.org/abs/astro-ph/9611174}{{\ttfamily astro-ph/9611174}}].

\bibitem{Bilbao-Ahedo:2021jhn}
J.D.~Bilbao-Ahedo, R.B.~Barreiro, P.~Vielva, E.~Mart\'\i{}nez-Gonz\'alez and D.~Herranz, \emph{{ECLIPSE: a fast Quadratic Maximum Likelihood estimator for CMB intensity and polarization power spectra}}, \href{https://doi.org/10.1088/1475-7516/2021/07/034}{\emph{JCAP} {\bfseries 07} (2021) 034} [\href{https://arxiv.org/abs/2104.08528}{{\ttfamily 2104.08528}}].

\bibitem{999016}
S.~Masnou and J.-M.~Morel, \emph{Level lines based disocclusion},  in \emph{Proceedings 1998 International Conference on Image Processing. ICIP98 (Cat. No.98CB36269)}, pp.~259--263 vol.3, 1998, \href{https://doi.org/10.1109/ICIP.1998.999016}{DOI}.

\bibitem{2008StMet...5..289A}
P.~{Abrial}, Y.~{Moudden}, J.L.~{Starck}, J.~{Fadili}, J.~{Delabrouille} and M.K.~{Nguyen}, \emph{{CMB data analysis and sparsity}}, \href{https://doi.org/10.1016/j.stamet.2007.11.005}{\emph{Statistical Methodology} {\bfseries 5} (2008) 289} [\href{https://arxiv.org/abs/0804.1295}{{\ttfamily 0804.1295}}].

\bibitem{2008PhRvD..77l3539I}
K.T.~{Inoue}, P.~{Cabella} and E.~{Komatsu}, \emph{{Harmonic inpainting of the cosmic microwave background sky: Formulation and error estimate}}, \href{https://doi.org/10.1103/PhysRevD.77.123539}{\emph{Phys. Rev. D} {\bfseries 77} (2008) 123539} [\href{https://arxiv.org/abs/0804.0527}{{\ttfamily 0804.0527}}].

\bibitem{2009arXiv0903.1308P}
L.~{Perotto}, J.~{Bobin}, S.~{Plaszczynski}, J.L.~{Starck} and A.~{Lavabre}, \emph{{Reconstruction of the CMB lensing for Planck}}, \href{https://doi.org/10.48550/arXiv.0903.1308}{\emph{arXiv e-prints} (2009) arXiv:0903.1308} [\href{https://arxiv.org/abs/0903.1308}{{\ttfamily 0903.1308}}].

\bibitem{2012MNRAS.424.1694B}
M.~{Bucher} and T.~{Louis}, \emph{{Filling in cosmic microwave background map missing data using constrained Gaussian realizations}}, \href{https://doi.org/10.1111/j.1365-2966.2012.21138.x}{\emph{MNRAS} {\bfseries 424} (2012) 1694} [\href{https://arxiv.org/abs/1109.0286}{{\ttfamily 1109.0286}}].

\bibitem{2012ApJ...750L...9K}
J.~{Kim}, P.~{Naselsky} and N.~{Mandolesi}, \emph{{Harmonic In-painting of Cosmic Microwave Background Sky by Constrained Gaussian Realization}}, \href{https://doi.org/10.1088/2041-8205/750/1/L9}{\emph{Astrophys. J. Lett.} {\bfseries 750} (2012) L9} [\href{https://arxiv.org/abs/1202.0188}{{\ttfamily 1202.0188}}].

\bibitem{1991ApJ...380L...5H}
Y.~{Hoffman} and E.~{Ribak}, \emph{{Constrained Realizations of Gaussian Fields: A Simple Algorithm}}, \href{https://doi.org/10.1086/186160}{\emph{Astrophys. J. Lett.} {\bfseries 380} (1991) L5}.

\bibitem{2013A&A...555A..37B}
A.~{Benoit-L{\'e}vy}, T.~{D{\'e}chelette}, K.~{Benabed}, J.F.~{Cardoso}, D.~{Hanson} and S.~{Prunet}, \emph{{Full-sky CMB lensing reconstruction in presence of sky-cuts}}, \href{https://doi.org/10.1051/0004-6361/201321048}{\emph{Astron. Astrophys.} {\bfseries 555} (2013) A37} [\href{https://arxiv.org/abs/1301.4145}{{\ttfamily 1301.4145}}].

\bibitem{Marcos-Caballero:2019jqj}
A.~Marcos-Caballero and E.~Mart\'\i{}nez-Gonz\'alez, \emph{{Scale-dependent dipolar modulation and the quadrupole-octopole alignment in the CMB temperature}}, \href{https://doi.org/10.1088/1475-7516/2019/10/053}{\emph{JCAP} {\bfseries 10} (2019) 053} [\href{https://arxiv.org/abs/1909.06093}{{\ttfamily 1909.06093}}].

\bibitem{LiteBIRD:2022cnt}
{\scshape LiteBIRD} collaboration, \emph{{Probing Cosmic Inflation with the LiteBIRD Cosmic Microwave Background Polarization Survey}}, \href{https://doi.org/10.1093/ptep/ptac150}{\emph{PTEP} {\bfseries 2023} (2023) 042F01} [\href{https://arxiv.org/abs/2202.02773}{{\ttfamily 2202.02773}}].

\bibitem{Puglisi:2020deh}
G.~Puglisi and X.~Bai, \emph{{Inpainting Galactic Foreground Intensity and Polarization Maps Using Convolutional Neural Networks}}, \href{https://doi.org/10.3847/1538-4357/abc47c}{\emph{Astrophys. J.} {\bfseries 905} (2020) 143} [\href{https://arxiv.org/abs/2003.13691}{{\ttfamily 2003.13691}}].

\bibitem{2020A&A...641A...5P}
{\scshape Planck} collaboration, \emph{{Planck 2018 results. V. CMB power spectra and likelihoods}}, \href{https://doi.org/10.1051/0004-6361/201936386}{\emph{Astron. Astrophys.} {\bfseries 641} (2020) A5} [\href{https://arxiv.org/abs/1907.12875}{{\ttfamily 1907.12875}}].

\bibitem{2017JCAP...02..022B}
J.D.~{Bilbao-Ahedo}, R.B.~{Barreiro}, D.~{Herranz}, P.~{Vielva} and E.~{Mart{\'\i}nez-Gonz{\'a}lez}, \emph{{On the regularity of the covariance matrix of a discretized scalar field on the sphere}}, \href{https://doi.org/10.1088/1475-7516/2017/02/022}{\emph{JCAP} {\bfseries 2017} (2017) 022} [\href{https://arxiv.org/abs/1701.06617}{{\ttfamily 1701.06617}}].

\bibitem{2011A&A...536A...1P}
{Planck Collaboration}, P.A.R.~{Ade}, N.~{Aghanim}, M.~{Arnaud}, M.~{Ashdown}, J.~{Aumont} et~al., \emph{{Planck early results. I. The Planck mission}}, \href{https://doi.org/10.1051/0004-6361/201116464}{\emph{Astron. Astrophys.} {\bfseries 536} (2011) A1} [\href{https://arxiv.org/abs/1101.2022}{{\ttfamily 1101.2022}}].

\bibitem{2020A&A...643A..42P}
{Planck Collaboration}, Y.~{Akrami}, K.J.~{Andersen}, M.~{Ashdown}, C.~{Baccigalupi}, M.~{Ballardini} et~al., \emph{{Planck intermediate results. LVII. Joint Planck LFI and HFI data processing}}, \href{https://doi.org/10.1051/0004-6361/202038073}{\emph{Astron. Astrophys.} {\bfseries 643} (2020) A42} [\href{https://arxiv.org/abs/2007.04997}{{\ttfamily 2007.04997}}].

\bibitem{Gorski:2004by}
K.M.~{G{\'o}rski}, E.~{Hivon}, A.J.~{Banday}, B.D.~{Wandelt}, F.K.~{Hansen}, M.~{Reinecke} et~al., \emph{{HEALPix: A Framework for High-Resolution Discretization and Fast Analysis of Data Distributed on the Sphere}}, \href{https://doi.org/10.1086/427976}{\emph{Astrophys. J.} {\bfseries 622} (2005) 759} [\href{https://arxiv.org/abs/astro-ph/0409513}{{\ttfamily astro-ph/0409513}}].

\bibitem{Zonca2019}
A.~Zonca, L.~Singer, D.~Lenz, M.~Reinecke, C.~Rosset, E.~Hivon et~al., \emph{healpy: equal area pixelization and spherical harmonics transforms for data on the sphere in python}, \href{https://doi.org/10.21105/joss.01298}{\emph{Journal of Open Source Software} {\bfseries 4} (2019) 1298}.

\bibitem{2020Natur.585..357H}
C.R.~{Harris}, K.J.~{Millman}, S.J.~{van der Walt}, R.~{Gommers}, P.~{Virtanen}, D.~{Cournapeau} et~al., \emph{{Array programming with NumPy}}, \href{https://doi.org/10.1038/s41586-020-2649-2}{\emph{Nature} {\bfseries 585} (2020) 357} [\href{https://arxiv.org/abs/2006.10256}{{\ttfamily 2006.10256}}].

\bibitem{2007CSE.....9...90H}
J.D.~{Hunter}, \emph{{Matplotlib: A 2D Graphics Environment}}, \href{https://doi.org/10.1109/MCSE.2007.55}{\emph{Computing in Science and Engineering} {\bfseries 9} (2007) 90}.

\bibitem{2020NatMe..17..261V}
P.~{Virtanen}, R.~{Gommers}, T.E.~{Oliphant}, M.~{Haberland}, T.~{Reddy}, D.~{Cournapeau} et~al., \emph{{SciPy 1.0: fundamental algorithms for scientific computing in Python}}, \href{https://doi.org/10.1038/s41592-019-0686-2}{\emph{Nature Methods} {\bfseries 17} (2020) 261} [\href{https://arxiv.org/abs/1907.10121}{{\ttfamily 1907.10121}}].

\end{thebibliography}\endgroup

\end{document}